\documentclass[11pt,xcolor=dvipsnames]{article}
\pdfoutput=1 
\usepackage[utf8x]{inputenc}
\usepackage[english]{babel}
\usepackage{graphicx}           
\usepackage{url}
\usepackage{eurosym}
\usepackage{makeidx}
\usepackage[plain]{algorithm}
\usepackage{algorithmic} 
\usepackage{color} 
\usepackage[margin=3.3cm]{geometry}
\usepackage{setspace}
\usepackage{datetime}
\usepackage{pifont}
\usepackage{tikz}
\usepackage{natbib}
\usepackage{hyperref}
\usepackage{amsmath,amsfonts,amssymb,latexsym,stmaryrd}

\newcommand{\eqdef}     {\stackrel{{\textrm{\rm\tiny def}}}{=}}

\newtheorem{theorem*}     {theorem}

\newsavebox{\fmbox}

\newcommand{\mmax}{{\textrm{max}}}

\newcommand{\N}        {\mathbb N}

\newcommand{\X}        {\mathbb X}

\newcommand{\R}      {\mathbb R}

\newcommand{\ot}        {\leftarrow}

\newcommand{\NN}   {{\mathcal N}}
\newcommand{\MM}   {{\mathcal M}}

\newcommand{\RR}   {{\mathcal R}}

\newcommand{\XX}   {{\mathcal X}}
\newcommand{\YY}   {{\mathcal Y}}

\newcommand{\rmd}   {{{\textrm{\upshape d}}}}

\newcommand{\dontforget}[1]
{{\mbox{}\\\noindent\rule{1cm}{2mm}\hfill  #1 \hfill\rule{1cm}{2mm}}\typeout{---------- #1 ------------}}

\renewcommand{\epsilon}{\varepsilon}

\def\dobm{
    \copy1\kern-\wd1\kern0.05ex\copy1\kern-\wd1\kern0.05ex\box1}

\newcommand
      {\sysdys}
      {{\sf S\kern-.15em\raise.3ex\hbox{Y}\kern-.15em
            SD\kern-.15em\raise.3ex\hbox{Y}\kern-.15emS}}

\newcommand{\normm}[1]{{\vert\kern-0.25ex\vert\kern-0.25ex\vert #1 
    \vert\kern-0.25ex\vert\kern-0.25ex\vert}}
\renewcommand{\X}{\mathcal{X}}

\renewcommand{\RR}{\mathbb{R}}
\renewcommand{\R}{\mathbb{R}}
\renewcommand{\NN}{\mathbb{N}}

\newcommand{\dif}{\mathrm{d}}

\newcommand{\mumax}{\mu_{\textrm{\tiny\rm max}}}
\newcommand{\Ks}{K_{\textrm{\tiny\rm s}}}
\renewcommand{\mmax}{{m_{\textrm{\tiny\rm max}}}}

\newcommand{\lambdab}{\bar\lambda}
\newcommand{\Exp}{{\textrm{\rm Exp}}}
\newcommand{\tmax}{t_{\textrm{\tiny\rm max}}}
\newcommand{\rmax}{r_{\textrm{\tiny\rm max}}}

\newcommand{\Sin}{{\mathbf s}_{\textrm{\tiny\rm in}}}
\newcommand{\mdiv}{m_{\textrm{\tiny\rm div}}}

\renewcommand{\eqdef}     {\stackrel{{\textrm{\rm\tiny def}}}{=}}

\usepackage{ifthen}    
\newboolean{showComments}        
\setboolean{showComments}{true}  
\newcommand{\fnote}[1]
    {{\mbox{}\\\noindent\color{red}
    \rule{1cm}{2mm}\hfill  #1 \hfill\rule{1cm}{2mm}}
    \typeout{---------- #1 ------------}}

\renewcommand{\dontforget}[1]
      {\ifthenelse {\boolean{showComments}} {{\color{red}(#1)}} {}}
\newcommand{\Dontforget}[1]
      {\ifthenelse {\boolean{showComments}} {\fnote{#1}} {}}

\begin{document}

\title{A modeling approach of the chemostat}
\author{
Coralie Fritsch\thanks{Montpellier 2 University and INRA/MIA\,, \texttt{Coralie.Fritsch@supagro.inra.fr}} \and 
Jérôme Harmand\thanks{INRA\,, \texttt{Jerome.Harmand@supagro.inra.fr}} \and 
Fabien Campillo\thanks{INRIA\,, \texttt{Fabien.Campillo@inria.fr}\protect\\ \emph{Coralie Fritsch and Fabien Campillo} are members of the MODEMIC joint INRA and INRIA project-team. MODEMIC Project-Team, INRA/INRIA, UMR MISTEA, 2 place Pierre Viala,
34060 Montpellier cedex 01, France. 
\emph{Jérôme Harmand} is member of the 
Laboratoire de Biotechnologies de l'Environnement, UR0050, INRA, Avenue des étangs, 11100 Narbonne, France}}

\maketitle

\begin{abstract}
Population dynamics and in particular microbial population dynamics, though they are complex but also intrinsically discrete and random, are conventionally represented as deterministic differential equations systems. We propose to revisit this approach by complementing these classic formalisms by stochastic formalisms and to explain the links between these representations in terms of mathematical analysis but also in terms of modeling and numerical simulations. We illustrate this approach on the model of chemostat.


\end{abstract}

\section{Introduction}
\label{sec.intro}

Biological continuous cultures in chemostat  play an important role in microbiology as well as in biotechnology. Different formulations are used to represent these processes. The mechanisms of growth and cell division may indeed be described at the cell level or at the population level. In the former case the mechanisms are discrete and random, usually represented as stochastic birth and death processes (BDP) or as stochastic individual-based models (IBM); in the latter case they are often supposed to be continuous and deterministic, and represented as systems of ordinary differential equations (ODE)  or as integro-differential equations (IDE) or partial derivative equations. The bridge from discrete/random to continuous/deterministic is achieved in the framework of a ``large population size'' asymptotic, that allows to prove, under certain assumptions, the convergence in distribution of the former models towards the latter ones \citep{campillo2014d}.

Hence in large population size, a simulation of the discrete/random model would be similar to that of the continuous/deterministic one. Of course this is true under certain assumptions, and especially in an asymptotic framework: in practice it is difficult to a priori know what large population size means. Beyond the mathematical analysis, it is possible to rely on numerical simulations to get an idea of the convergence of these former models to the latter ones. It is also interesting to understand how the former models behave when they are not close to the latter, that is to say when the continuous/deterministic models are no longer valid.

Beyond the antagonism  discrete/random \emph{vs} continuous/deterministic, it seems appropriate to propose a new modeling approach where the ``model'' is not a specific computer or mathematical representation defined once and for all but rather a set of representations and to infer the links between these representations, the scope of validity of each different representations, as well as the capabilities of the associated simulation and control tools.

The first model of the chemostat appeared in the 50's  \citep{monod1950a,novick1950a}. This first model has always retained its relevance in particular because of its simplicity \citep{smith1995a}. Several other models have appeared later like the so-called population balance models proposed by \cite {fredrickson1967a}  that rely on a representation of the population  structured in mass \citep{ramkrishna1979a}.

More recently several stochastic models in (unstructured) population size  appeared in order to  account for the demographic or environmental sources of randomness \citep{crump1979a,stephanopoulos1979a,imhof2005a,grasman2005a,campillo2011chemostat}. In particular, for the demographic noise, according to a now classic approach, the model described at the level of the individual is a discrete stochastic birth and death process that can be approximated at a meso-scale by a continuous diffusion process when population sizes are large enough, and that reduces at a macro-scale to the solution of the classic chemostat ODE when these population sizes are very large.

The individual-based model that we propose in this paper has been studied mathematically in \cite{campillo2014d} where we proved  in particular its convergence in distribution to the solution of an IDE similar to that proposed in \cite{fredrickson1967a}. 

\bigskip

We propose in this article to illustrate this approach on the model of chemostat: starting from the classical ordinary differential equation model in dimension 2, we propose other representations in the form of an integro-differential equation (continuous and deterministic) or as an individual-based model (discrete and random) both structured in mass. By model reduction, these representations can be reduced to the classical model (continuous and deterministic) or as a birth and death process (discrete and random). We explain the links between these different representations of the same model, as well as their respective advantages and limitations, specifically in terms of simulation.

In Section \ref{sec.model} we introduce the different models, we detail in particular the proposed IBM. In Section \ref{sec.algo} we describe the (almost) exact simulation algorithm of the IBM. 
Using simulations, in Section \ref{sec.simulations} we highlight the differences between each of these representations. The paper ends with a discussion in Section 
\ref{sec.discussion}.

\section{The models}
\label{sec.model}

\subsection{The ODE model}

The classic chemostat model reads:
\begin{align}
\label{eq.chemostat.edo.1}
	\dot S_t	& = D\,(\Sin-S_t)-k \,\mu(S_t)\, Y_t
	\\
\label{eq.chemostat.edo.2}
	\dot Y_t 	& = \bigl(\mu(S_t)-D \bigr)\, Y_t\,,
\end{align}
where $S_{t}$ and $Y_{t}$ are respectively the \emph{substrate concentration} and the \emph{bacterial concentration} (mg/l) which are assumed to be uniform in the vessel; $D$ is the dilution rate (1/h), $\Sin$ is the substrate input concentration (mg/l), $k$ is the (inverse of) yield constant.
The specific growth rate $\mu$ could for example be the classic Monod kinetics:
\begin{align}
\label{eq.edo.monod}
	\mu(s)
	& = 
	\mumax \, \frac{s}{\Ks+s}
\end{align}
with maximum specific growth rate $\mumax$ and half-velocity constant $\Ks$.

In biochemical engineering, System \eqref{eq.chemostat.edo.1}-\eqref{eq.chemostat.edo.2} corresponds to the classic continuous stirred-tank reactor (CSTR) under well-mixing conditions \citep{smith1995a}.

\subsection{The IDE model}

Instead of representing the dynamic of the bacterial population inside the chemostat through the aggregated state variable $Y_{t}$, one may wish to represent the state of the bacterial population structured in mass, that is to consider the density of population $p_{t}(x)$ w.r.t. their mass in a reference volume $V$. Hence $\int_{m_{0}}^{m_{1}} p_{t}(x)\,\rmd x$ is the number of cells which mass is between $m_{0}$ and $m_{1}$ and the link with the bacterial concentration is:
\[
  Y_t \eqdef \frac1V \,\int_0^\mmax x\, p_t(x) \, \dif x
\] 
where $0<\mmax<\infty$ is an upper bound for the mass of a bacterium.
The evolution equation for the couple $(S_{t},p_{t}(x))$ has been established by \cite{fredrickson1967a} as the population balance equations for growth-fragmentation models \citep[see also][]{ramkrishna1979a}, they read:
\begin{align} 
\label{eq.limite.substrat.fort}
	&
	\dot S_t  = 
	D\,(\Sin-S_t)-\frac kV \int_0^\mmax \rho(S_t,x)\,p_t(x)\,\dif x\,,
\\
\nonumber
	&
	\frac{\partial}{\partial t} p_t(x)
	+\frac{\partial}{\partial x} \bigl( \rho(S_t,x)\,p_t(x)\bigr)
	+ \bigl(\lambda(S_t, x)+D \bigr)\,p_t(x)
\\
\label{eq.limite.eid.fort}
  	&\qquad\qquad\qquad\qquad\qquad\qquad\qquad\qquad
	=  
	2\,\int_0^\mmax 
	\frac{\lambda(S_t, z)}{z}\,q\left(\frac xz \right)\,p_t(z)\,\dif z
\end{align}
for $x\in[0,\mmax]$. Here, like in the previous model $S_{t}$ is the substrate concentration (mg/l) which is assumed to be uniform in the vessel.

In \eqref{eq.limite.substrat.fort}-\eqref{eq.limite.eid.fort},
 $\rho(s,x)$ and $\lambda(s,x)$ are respectively the growth function and the division rate of a bacterium of mass $x$ with a substrate concentration $s$, the mass distribution of the daughter cells is represented by the probability density function   $q(\alpha)$ on $[0,1]$. We detail these functions now:

\begin{enumerate}

\item{\textbf{Cell division}} --
Each individual of mass $x$ divides itself at a rate $\lambda(s,x)$ into two
individuals with respective masses  $\alpha\,x$ and $(1-\alpha)\,x$:
\begin{center}
\includegraphics[width=5cm]{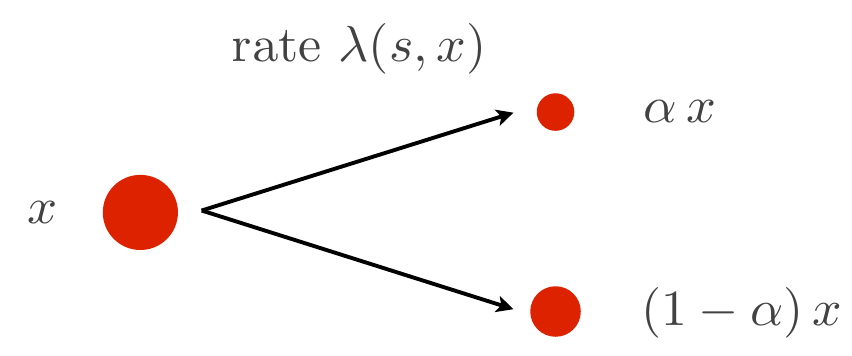}
\end{center}
where $\alpha$ is distributed according to a given probability density function   $q(\alpha)$ on $[0,1]$,
and  $s$ is the substrate concentration.
We suppose that the p.d.f.  $q(\alpha)$ is symmetric with respect to $\frac{1}{2}$, i.e. $q(\alpha)=q(1-\alpha)$:
\begin{center}
\includegraphics[width=4cm]{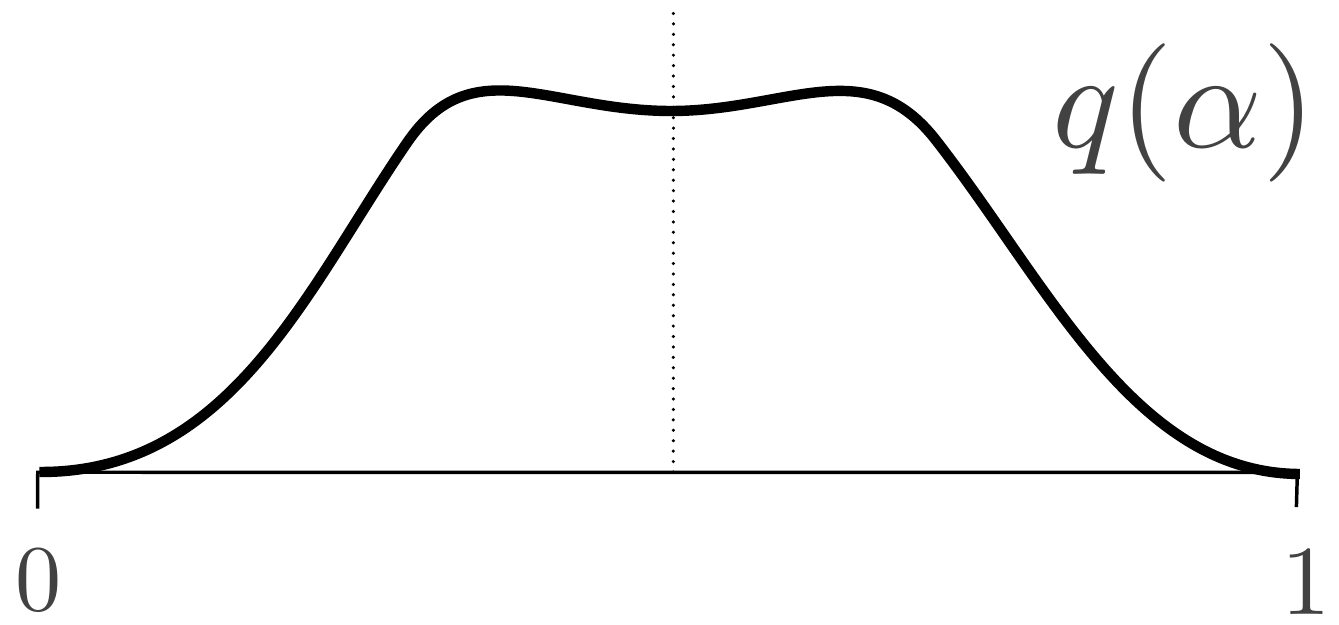}
\end{center}
Hence, the  p.d.f. of the division kernel of a cell of mass $x$ is $q(y/x)$ with support $[0,x]$. In the case of perfect mitosis, a cell of mass $x$ is divided into two cells of masses $\frac x 2$ so that $q(\alpha)=\delta_{1/2}(\alpha)$. We suppose that $q$ is smooth (which is not the case for the perfect mitosis) and that $q(0)=q(1)=0$.
Thus, relatively to their mass, the division kernel is the same for all individuals. This allows us to reduce the model to a single division kernel but more complex possibilities  can also be investigated.

\item{\textbf{Mass growth}} --
The growth function $\rho:\R_{+}\times[0,\mmax]\mapsto\R_{+}$  describes the evolution of the mass of an individual cell within the chemostat, i.e.  in the model  \eqref{eq.limite.substrat.fort}-\eqref{eq.limite.eid.fort} the mass of an individual cell starting from the mass $m_{0}$ at a given time $t_{0}$ will evolve according to: 
\begin{align*}
   \dot x_{t} 
   =
   \rho(S_{t},x_{t})\,,\quad
   t\geq t_{0}\,,\ x_{0}=m_{0}
\end{align*}
until the time of division or uptake.
To ensure the existence and uniqueness of the solution of \eqref{eq.limite.substrat.fort}-\eqref{eq.limite.eid.fort} and of this last EDO, we assume
that application $\rho(s,x)$ is Lipschitz continuous w.r.t. $s$ uniformly in $x$. To ensure a coherence to that equations we also suppose that 
$  0
  \leq
  \rho(s,x)
  \leq 
  \bar \rho
$
for all $(s,x)\in\RR_+ \times [0,\mmax]$, and that 
 in the absence of substrate the bacteria do not grow, i.e.
$  \rho(0,x)
  =
  0
$
for all $x\in [0,\mmax]$. To ensure that the mass of a bacterium  
stays between $0$ and $\mmax$, it is finally assumed that
$  \rho(s,\mmax)
  =
  0
$
for any $s\geq 0$.

\end{enumerate}

In a relaxed context where $\rho(s,x)$ does not satisfy the previous hypothesis,
it is easy to link  the model \eqref{eq.limite.substrat.fort}-\eqref{eq.limite.eid.fort}
to the classic chemostat model \eqref{eq.chemostat.edo.1}-\eqref{eq.chemostat.edo.2}. Indeed suppose that:
\[
  \frac1V \int_\X  \rho(S_t,x)\,p_t(x) \, \dif x
	= \mu(S_{t})\,Y_{t}
\]
which is the case when  the growth function $x \mapsto \rho(s,x)$ is proportional to $ x $, i.e.  $ \rho(s,x)=\mu(s)\,x$. First  \eqref{eq.limite.substrat.fort} reduces to \eqref{eq.chemostat.edo.1} and then we can check that $Y_{t}$ is is solution of \eqref{eq.chemostat.edo.2} \citep[see details in][]{campillo2014d}.

\subsection{The BDP model}

We consider an hybrid model, where the substrat concentration $S_{t}$ follows the same  continuous/deterministic dynamic \eqref{eq.chemostat.edo.1}:
\begin{align}
\label{eq.bdp.s}
	\dot S_t	& = D\,(\Sin-S_t)-k \,\mu(S_t)\, \frac{m}{V}\,\YY_t
\end{align}
but now  $m$ is the mean mass of an individual cell and $\YY_{t}$ is the number of cells in the chemostat. The dynamic of $\YY_{t}$ is discrete/stochastic, namely a birth and death stochastic process where at time $t$ and conditionally to $\YY_{t}=n$, the process jumps from $n$ to $n+1$ with rate $\mu(S_{t})$ and jumps from $n$ to $n-1$ with rate $D$, that is:
\begin{align}
\label{eq.bdp.y}
  \YY_{t+h}
  =
  n
  +
  \begin{cases}
    1  & \textrm{with probability }\mu(S_{t})\,n\,h+o(h)\,,\\
    -1  & \textrm{with probability }D\,n\,h+o(h)\,,\\
    0  & \textrm{with probability }1-\mu(S_{t})\,n\,h-D\,n\,h+o(h)\,,\\
    i  & \textrm{with probability } o(h) \textrm{ for all }i\neq 0,1,-1
  \end{cases}
\end{align}
for infinitesimally small $h>0$.

\subsection{The IBM model}

In the \emph{individual-based model (IBM) structured in mass} the
bacterial population is represented as a set of individuals growing in a perfectly mixed vessel of volume $V$ (l). Each individual is characterized only by its mass $x\in [0,\mmax]$. At time $t$ the state of the system is defined by:
\begin{align}
\label{eq.xi}
  (S_{t},\nu_{t})
\end{align}
where $S_{t}$ is the \emph{substrate concentration} (mg/l) which is supposed to be uniform in the vessel; and $\nu_{t}$ will represent the state of  the \emph{bacterial population}, that is $N_{t}$ individuals
represented only by their mass:  $x^i_{t}$ (mg)  will denote the mass of the individual number $i$ for $i=1,\dots,N_{t}$. 

It will be convenient to represent the population $\{x^i_{t}\}_{i=1,\dots,N_{t}}$ at time $t$ as the following sum of Dirac delta functions:
\begin{align}
\label{eq.nu}
  \nu_t(x)=\sum_{i=1}^{N_t}\delta_{x_t^i}(x)\,.
\end{align}
where $\delta_{x_t^i}(x)$ is the Dirac delta function in $x_t^i$: $\int \phi(x)\,\delta_{x_t^i}(x)\,\rmd x=\phi(x_t^i)$ for any test function $\phi$. For example 
   $\int_{m_{0}}^{m_{1}} \nu_t(x)\,\rmd x$
 is number of cells with mass between $m_{0}$ and $m_{1}$ at time $t$;
 and $\int_{m_{0}}^{m_{1}} x\,\nu_t(x)\,\rmd x$
is the cumulated  mass of cells with mass between $m_{0}$ and $m_{1}$ at time $t$
\citep[see][for more details on this representation]{dieckmann2000b}.

The IBM dynamic combines discrete evolutions (cell division and bacterial up-take) as well as continuous evolutions (the growth of each individual and the dynamic of the substrate). We now describe the four components of the dynamic, first the discrete ones and then the continuous ones which occur between the discrete ones.

\begin{enumerate}
\vskip0.8em
\item{\textbf{Cell division}} --
\emph{Each individual of mass $x$ divides at rate $\lambda(s,x)$ into two
individuals of respective masses  $\alpha\,x$ and $(1-\alpha)\,x$
where $\alpha$ is distributed according to the given p.d.f.  $q(\alpha)$ on $[0,1]$,
and  $s$ is the substrate concentration.}

\vskip0.8em
\item{\textbf{Up-take}} --
\emph{Each individual is withdrawn from the chemostat at rate~$D$.}
This mechanism is equivalent to a death process. In \emph{perfect mixing} conditions, individuals are uniformly distributed in the volume $V$ independently from their mass. During a time step $\delta$, a total volume of $D\,V\,\delta$ is withdrawn from the chemostat:
\begin{center}
\includegraphics[width=10cm]{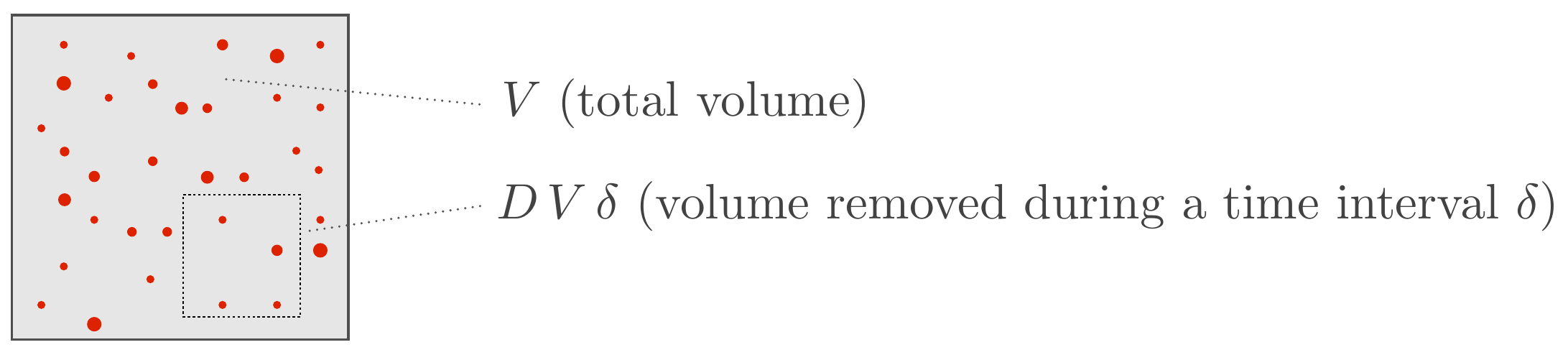}
\end{center}
and therefore, if we assume that all individuals have the same volume considered as negligible, during this time interval $\delta$, an individual has a probability $D\,\delta$ to be withdrawn from the chemostat, $D$ is the dilution rate. This rate could possibly depend on the mass of the individual.

\end{enumerate}
At any time $t$, when the division of an individual occurs, the size of the population instantaneously jumps from $N_{t}$ to $N_{t}+1$; when an individual is withdrawn from the vessel, the size of the population jumps instantaneously from $N_{t}$ to $N_{t}-1$; between each discrete event the size $N_{t}$ remains constant and the chemostat evolves according to the following two continuous mechanisms:
\begin{enumerate}
\setcounter{enumi}{2}
\vskip0.8em
\item {\textbf{Growth of each cell}} --
\emph{Each cell of mass $x$ growths at speed $\rho(S_{t},x)$}:
\begin{align}
\label{eq.masses}
   \dot x^i_{t} 
   =
   \rho(S_{t},x^i_{t})\,, \quad i=1,\dots,N_{t}
\end{align}
where $\rho:\R^2_{+}\mapsto\R_{+}$ is given. 

\item{\textbf{Dynamic of the substrate concentration}} -- 
\emph{The substrate concentration evolves according to the ODE:
\begin{align}
\label{eq.substrat}
  \dot S_t = D\,(\Sin-S_t)-k\, {\tilde\mu}(S_t,\nu_{t})
\end{align}}
where
\begin{align*}
  {\tilde\mu}(s,\nu)
  &\eqdef
  \frac{1}{V} \int_{\XX} \rho(s,x)\,\nu(\rmd x)
  =    
  \frac{1}{V} \sum_{i=1}^{N} \rho(s,x^i)
\end{align*}
with $\nu=\sum_{i=1}^N \delta_{x^i}$.
Mass balance leads to Equation \eqref{eq.substrat} and the initial condition $S_{0}$ may be random.

\end{enumerate}
The IBM integrate the function $\lambda(s,x)$, $q(\alpha)$ and $\rho(s,x)$ already defined in the IDE but in a different way: the IDE uses them in an ``average way'' at the population level in contrast with the IBM that uses them  for the explicit dynamic of each individual cell.

\subsection*{Convergence in distribution of the individual-based model}

\cite{campillo2014d} proved a result that we will comment now on the application point of view. This result states a ``functional'' law of large numbers: in large population size the density of population given by the IBM is close to the density of population $p_{t}(x)$ given by \eqref{eq.limite.eid.fort}. The population size should increase to infinity at any time $t$, for that purpose we replace the reference volume $V$ by $ n\,V$ (or simply by $n$), let:
\[
  V_{n} = n\,V
\]
We also suppose that the initial population size converges toward infinity with $n$:
\begin{align*}
   \frac{1}{n}\nu_0^n
   \xrightarrow[n\to\infty]{} 
   \xi_0 
   \textrm{ weakly }
\end{align*}
that is $\int_{0}^\mmax \phi(x)\,\nu_0^n(x)\,\rmd x 
= \frac1n\sum_{i=1}^{N_{0}^n}\phi(x^{i,n}_{0})
\to \int_{0}^\mmax \phi(x)\,\xi_0(x)\,\rmd x$ as $n\to\infty$, and we suppose that 
$\int_{0}^\mmax\xi_0(x)\,\rmd x>0$. We suppose that the initial substrate concentration does not depend on $n$. 
Then define $(S^n_t,\nu_t^n(x))$ the IBM process where $V$ is replaced by $V_{n}$ and the rescaled process:
\begin{align*}
   \bar\nu_t^n
   \eqdef
   \frac{1}{n}\nu_t^n\,.
\end{align*}
Under these conditions \cite{campillo2014d} stated that the process $(S^n_{t},\bar\nu_t^n)_{0\leq t\leq T}$ given by the IBM converges toward the solution $(S_{t},\bar p_t)_{0\leq t\leq T}$ of the IDE model \eqref{eq.limite.substrat.fort}-\eqref{eq.limite.eid.fort} in a suitable sense with initial condition $(S_{0}, \xi_{0})$.

\section{Simulation of the models}
\label{sec.algo}

The simulation of the ODE system \eqref{eq.chemostat.edo.1}-\eqref{eq.chemostat.edo.2} is straightforward; for the simulation of the IDE system \eqref{eq.limite.substrat.fort}-\eqref{eq.limite.eid.fort} we make use of an explicit Euler time-scheme coupled with an uncentered upwind finite difference space-scheme 
(see details in Appendix \ref{appendix.schema.num}).

\subsection{Simulation of BDP model}

The simulation of the system \eqref{eq.bdp.s}-\eqref{eq.bdp.y} is achieved with an adaptation of the classic ``stochastic simulation algorithm'' (SSA) also called ``Gillespie algorithm'' \citep{gillespie1977a}. It is an exact simulation algorithm, up to the approximation scheme for the ODE \eqref{eq.bdp.s}, in the sense that it simulates a realization of the exact distribution of the stochastic process $(S_{t},\YY_{t})$ given by \eqref{eq.bdp.s}-\eqref{eq.bdp.y}. To apply the algorithm we need to suppose that there exists $\bar\mu<\infty$ such that:
\[ 
   \mu(s)\leq \bar\mu\,,\ \forall s\geq 0\,.
\]
Then the SSA is given by the Algorithm \ref{algo.ssa}.

\begin{algorithm}
\begin{center}
\begin{minipage}{14cm}
\small
\begin{algorithmic}
\STATE sample $(S_0,\YY_0)$
\STATE $\YY \ot \YY_0$
\STATE $t\ot 0$
\WHILE {$t\leq \tmax$}
  \STATE $\tau \ot (\bar\mu+D)\,\YY$
  \STATE $\Delta t \sim \Exp(\tau)$
  \STATE integrate the equation for substrate \eqref{eq.bdp.s} over $[t,t+\Delta t]$
  \STATE $t \ot t+\Delta t$
  \STATE $u\sim U[0,1]$ 
  \IF {$u\leq \mu(S_{t})/(\bar\mu+D)$}
      \STATE $\YY \ot \YY+1$
      \COMMENT{division}
  \ELSIF{$u\leq (\mu(S_{t})+D)/(\bar\mu+D)$}
      \STATE $\YY \ot \YY-1$
      \COMMENT{up-take}
  \ENDIF
\ENDWHILE
\end{algorithmic}
\end{minipage}
\end{center}
\caption{\itshape Stochastic simulation algorithm (SSA) or Gillespie algorithm for the Monte Carlo simulation of the BDP model \eqref{eq.bdp.s}-\eqref{eq.bdp.y}.}
\label{algo.ssa}
\end{algorithm}

\subsection{Simulation of the IBM}

We now detail the simulation procedure of the IBM. The division rate $\lambda(s,x)$ depends on the concentration of substrate $s$ and on the mass  $x$ of each individual cell which continuously evolves according to the system  \eqref{eq.masses}-\eqref{eq.substrat}, so to simulate the division of the cell we make use of a rejection sampling technique. It is assumed that there exists $\lambdab<\infty$ such that:
\[
  \lambda(s,x)\leq \lambdab
\]
hence an upper bound for the rate of event, division and up-take combined, at the population level is given by:
\[
    \tau \eqdef (\bar\lambda+D)\,N\,.
\]
\begin{algorithm}
\begin{center}
\begin{minipage}{14cm}
\small
\begin{algorithmic}
\STATE sample $(S_0,\nu_0=\sum_{i=1}^{N_{0}}\delta_{x^i_{t}})$
 \COMMENT{initial substrate concentration and population}
\STATE $t\ot 0$
\STATE $N\ot N_{0}$ \COMMENT{initial population size}
\WHILE {$t\leq \tmax$}
  \STATE $\tau \ot (\bar\lambda+D)\,N$
  \STATE $\Delta t \sim \Exp(\tau)$
  \STATE integrate the equations for the mass \eqref{eq.masses} 
    and the substrate \eqref{eq.substrat} over $[t,t+\Delta t]$
  \STATE $t \ot t+\Delta t$
  \STATE draw $x$ uniformly in $\{x^i_{t}\,;\,i=1,\dots,N_{t}\}$  
  \STATE $u\sim U[0,1]$ 
  \IF {$u\leq \lambda(S_{t},x)/(\lambdab+D)$}
      \STATE $\alpha \sim q$
      \STATE $\nu_{t} \ot \nu_{t} -\delta_{x}+\delta_{\alpha\,x}
              +\delta_{(1-\alpha)\,x}$
      \COMMENT{division}
      \STATE $N\ot N+1$
  \ELSIF{$u\leq (\lambda(S_{t},x)+D)/(\lambdab+D)$}
      \STATE $\nu_{t} \ot \nu_{t} -\delta_{x}$
      \COMMENT{up-take}
      \STATE $N\ot N-1$
  \ENDIF
\ENDWHILE
\end{algorithmic}
\end{minipage}
\end{center}
\caption{\itshape ``Exact'' Monte Carlo  simulation of the individual-based model:
approximations only lie in the numerical integration of the ODEs and in 
the pseudo-random numbers generators.}
\label{algo.ibm}
\end{algorithm}
At time $t+\Delta t$ with $\Delta t\sim \Exp(\tau)$, we determine if an event has occurred and what is its type by acceptance/rejection.
To this end, the masses of the $N$ individuals and the substrate concentration evolve according to the coupled ODEs \eqref{eq.masses} and \eqref{eq.substrat}.
Then we choose uniformly at random an individual within the population $\nu_{(t+\Delta t)^-}$, that is the population at time  $t+\Delta t$ before any possible event, let $x_{(t+\Delta t)^-}$ denotes its mass, then:
\begin{enumerate}

\item With probability:
\[
   \frac{\bar\lambda}{(\bar\lambda+D)}
\]
we determine if there has been division by acceptance/rejection:
\begin{itemize}
\item division occurs, that is:
\begin{align}
\label{eq.event.division}
   \nu_{t+\Delta t}
   =
   \nu_{(t+\Delta t)^-}
   -\delta_{x_{(t+\Delta t)^-}}  
   +\delta_{\alpha\,x_{(t+\Delta t)^-}}  
   +\delta_{(1-\alpha)\,x_{(t+\Delta t)^-}}  
   \qquad
   \textrm{with }\alpha\sim q
\end{align}
with probability $\lambda(S_{t},x_{(t+\Delta t)^-})/\bar\lambda$;
\item 
no event occurs with probability $1-\lambda(S_{t},x_{(t+\Delta t)^-})/\bar\lambda$.
\end{itemize}
In conclusion, the event \eqref{eq.event.division} occurs with probability:
\[
  \frac{\lambda\bigl(S_{t},x_{(t+\Delta t)^-}\bigr)}{\bar\lambda}
  \,\frac{\bar\lambda}{(\bar\lambda+D)} 
  = 
  \frac{\lambda\bigl(S_{t},x_{(t+\Delta t)^-}\bigr)}{(\bar\lambda+D)}\,.
\]

\item With probability:
\[
   \frac{D}{(\bar\lambda+D)}
   =
   1-\frac{\bar\lambda}{(\bar\lambda+D)}
\]
the individual is withdrawn, that is:
\begin{align}
\label{eq.event.soutirage}
   \nu_{t+\Delta t}
   =
   \nu_{(t+\Delta t)^-}
   -\delta_{x_{(t+\Delta t)^-}}  
\end{align}
\end{enumerate}
Finally, the events and the associated probabilities are:
\begin{itemize}
\item 
division \eqref{eq.event.division} with probability
$\lambda(S_{t},x_{(t+\Delta t)^-})/(\bar\lambda+D)$,
\item
up-take \eqref{eq.event.soutirage} with probability 
${D}/{(\bar\lambda+D)}$
\end{itemize}
and no event (rejection) with the remaining probability.
The details are given in Algorithm~\ref{algo.ibm}.

Technically, the numbering of individuals is as follows: at the initial time individuals are numbered from $1$ to $N$, in case division the daughter cell $\alpha \,x$ keeps the index of the parent cell and the daughter cell $(1-\alpha)\,x$ takes the index $N+1$; in case of the up-take, the individual $N$ acquires  the index of the withdrawn cell.

\section{Simulation tests}
\label{sec.simulations}

We present simulations of four different models : the individual based-model (IBM), the integro-differential equation (IDE) \eqref{eq.limite.substrat.fort}-\eqref{eq.limite.eid.fort}, the classic chemostat model represented by the system of ordinary differential equations (ODE) \eqref{eq.chemostat.edo.1}-\eqref{eq.chemostat.edo.2}, and the birth and death process (BDP) \eqref{eq.bdp.s}-\eqref{eq.bdp.y}. This models can have similar or different behaviors, depending on the model parameters and initial conditions.

Simulations of the BDP and of the IBM were performed respectively by Algorithms \ref{algo.ssa} and \ref{algo.ibm}.
The resolution of the integro-differential equation was made following the numerical scheme given in Appendix \ref{appendix.schema.num}, with a discretization step in the mass space  of $\Delta x = 2 \times 10^{-7}$ and a discretization step in time of  $\Delta t = 0.00125$.
The numerical integration of the ODE \eqref{eq.chemostat.edo.1}-\eqref{eq.chemostat.edo.2} presents no difficulties
and is performed by the function \texttt{odeint} of the module  \texttt{scipy.integrate} of \texttt{Python} with the default parameters.

\subsection{Simulation parameters}

For simulation purpose,
at fixed substrate concentration, individual growth is supposed to be given by a Gompertz function. Moreover we assume that the specific growth rate of the population depends on the substrate concentration and follows a Monod law: 
\begin{align}
\label{eq.g}
	\rho(s,x)
	= 
	\rmax\,\frac{s}{k_r+s} \,
	\log\Big(\frac{\mmax}{x}\Big)\,x
	\leq 
	\rmax\,\log\Big(\frac{\mmax}{x}\Big)\,x\,
\end{align}
where $\rmax$ is the maximum specific growth rate of the population, $k_r$ is the half-saturation constant and $\mmax$ is the maximal size of individual. 

\begin{figure}
\begin{center}
\includegraphics[width=5cm]{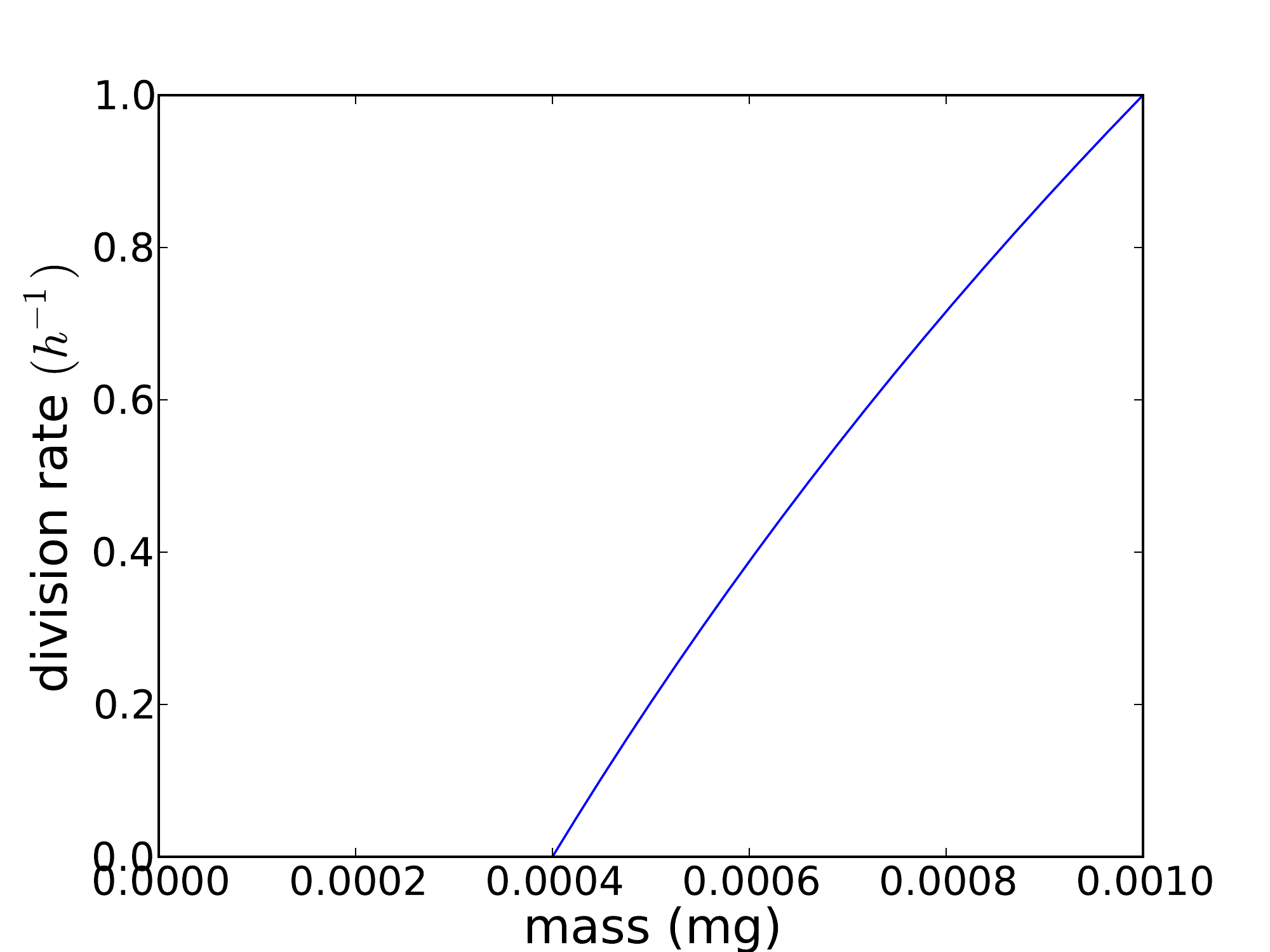}
\includegraphics[width=5cm]{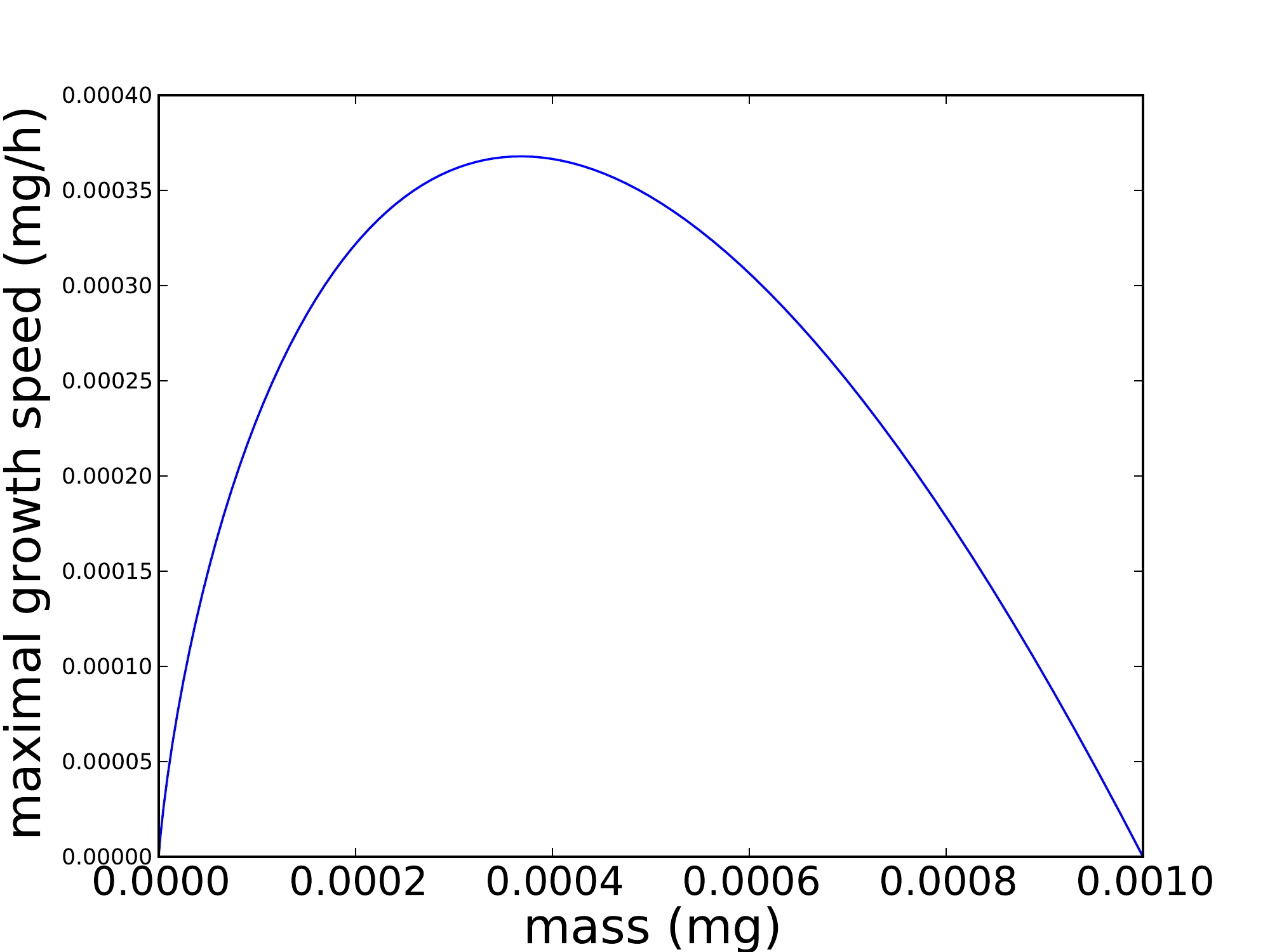}
\end{center}
\caption{$\blacktriangleright$ (Left) Division rate function $\lambda(x)$ defined by \eqref{eq.lambda} with $\bar\lambda = 1$ h$^{-1}$, $\mdiv=0.0004$ mg and $p_\lambda = 1000$. $\blacktriangleright$ (Right) Maximal growth speed with the Gompertz growth speed function \eqref{eq.g} with $\rmax=1.0$ h$^{-1}$, $\mmax=0.001$ mg (namely the RHS of the inequality \eqref{eq.g}).}
\label{fig.parameters}
\end{figure}

We assume that one individual can not divide below a mass $\mdiv$. For the simulations, we choose the following increasing division rate function :
\begin{align}
\label{eq.lambda}
	\lambda(s,x)=\lambda(x)
	=
	\frac{\bar\lambda}{\log \bigl((\mmax-\mdiv) \, p_\lambda +1 \bigr)} 
	\,
	\log \bigl((x-\mdiv) \, p_\lambda +1 \bigr) \, 1_{\{x \geq \mdiv\}}
\end{align}
where $p_\lambda>0$ is a parameter of curvature of the function, see Figure \ref{fig.parameters}. This ``ad hoc'' function has been chosen as it meets the desired conditions.

The proportion $\alpha$ of the division kernel $q(\alpha)$ will be computed by a symmetric beta distribution:
\begin{align*}
	q(\alpha)
	& = 
	\frac{1}{B(p_\beta)}\,
		\bigl( \alpha \,(1-\alpha) \bigr)^{p_\beta-1}
\end{align*}
where $B(p_\beta)
	= \int_0^1 \bigl( \alpha \,(1-\alpha) \bigr)^{p_\beta-1} \, \dif \alpha$
is a normalizing constant.

The initial distribution of individual masses is given by the following probability density function:
\begin{align}
\label{eq.d}
d(x)
	& =	
	    \frac{1}{C_d} \,
		\Biggl(
			\frac{x-0.0005}{0.00025}
			\,\left(1-\frac{x-0.0005}{0.00025}\right)
		\Biggr)^5 \,
		1_{\{0.0005 < x < 0.00075\}}\,
\end{align}
where $C_d$ is a normalizing constant.
This initial density will show a transient phenomenon that cannot be reproduced by the classic chemostat model described in terms of ordinary differential equations \eqref{eq.chemostat.edo.1}-\eqref{eq.chemostat.edo.2}, see Figure \ref{fig.edo.ibm.eid}.

\subsection{Comparison of the IBM and the IDE}

In this section we are going to illustrate the convergence of the IBM to the IDE. For that, we increase the volume of the chemostat and the initial number of individuals in a proportional way. 
We realize simulations at three levels of population size. The small size level is performed with $V=0.05$ l and $N_0=100$, the medium one with $V=0.5$ l and $N_0=1000$ and the large one with $V=5$ l and $N_0=10000$. The initial distributions of individual masses are the same, so that the initial biomass concentration is the same for the three sets of parameters.

For each level of population size, we simulate 100 independent runs of the IBM in order to observe the reduction of variance when we increase the number of bacteria.
The IDE is numerically approximated using the finite difference schemes detailed in Appendix \ref{appendix.schema.num}.
The parameters are given in the Table \ref{table.parametres}.

\begin{table}[h]
\begin{center}
\begin{tabular}{|c|c|}
	\hline
    Parameters & Values \\
    \hline
    $S_0$				&	6 mg/l \\
    $\Sin$				&	10 mg/l \\
    $D$					&	0.25 h$^{-1}$\\
    $\mmax$				&	0.001 mg \\	
    $\mdiv$				&	0.00045 mg \\
	$\bar\lambda$		&	1.5 h$^{-1}$\\
    $p_\lambda$			&	600 \\
    $p_\beta$			&	10 \\
    $\rmax$				&	1 h$^{-1}$\\
    $k_r$				&	6 mg/l\\
    $k$					&	1\\
    \hline
\end{tabular}
\end{center}
\caption{Simulation parameters.}
\label{table.parametres}
\end{table}

\bigskip

\begin{figure}
\begin{center}
\includegraphics[width=14cm]{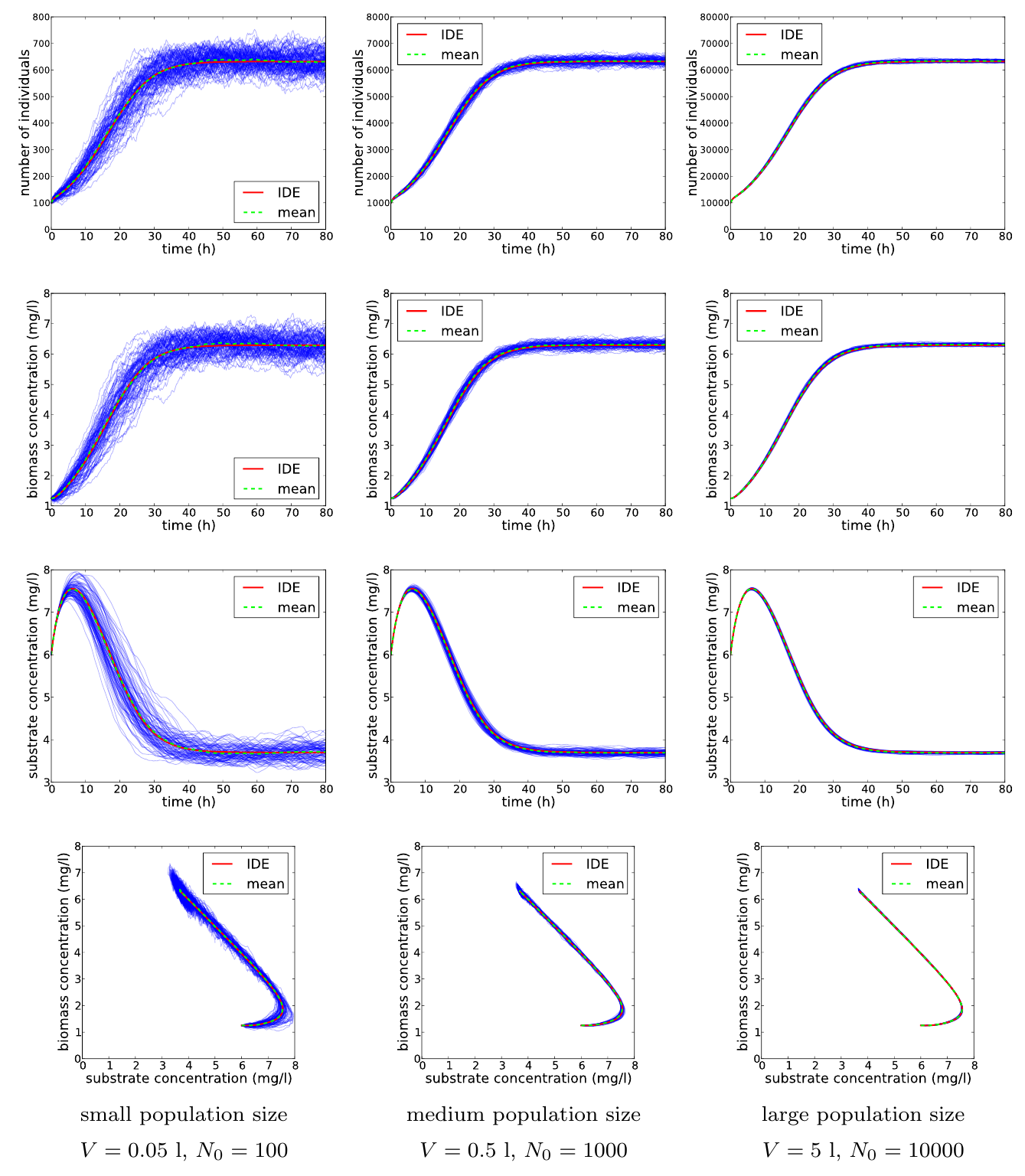}
\end{center}
\vskip-1em
\caption{From top to bottom: time evolutions of the population size, the biomass concentration, the concentration substrate and  the concentrations phase portrait  for the three levels of population sizes (from left to right: small, medium and large). The blue curves are the trajectories of 100 independent runs of IBM. The green curve is the mean value of these runs. The red curve is the solution of the IDE.}
\label{fig.evol.taille.concentrations}
\end{figure}

\begin{figure}
\begin{center}
\includegraphics[width=14cm]{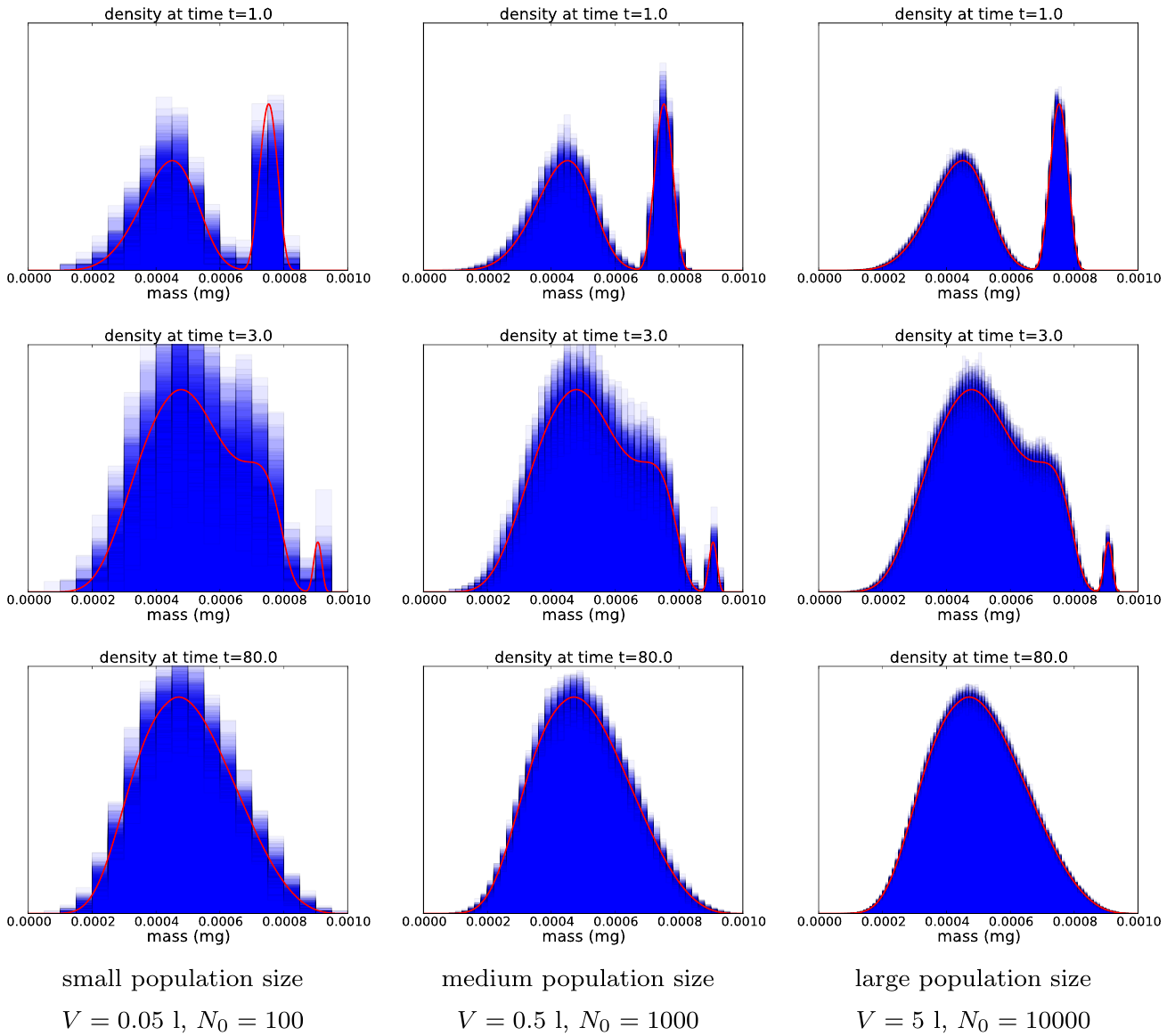}
\end{center}
\caption{Mass distribution for the time $t=1$ (top), $t=3$ (middle) and $t=80$ (bottom) in small (left), medium (middle) and large (right) population size. For each graph, the blue histograms represent the empirical mass distributions of individuals for the 100 independent runs of IBM. In order to plot the histogram we have adapted the number of bins according to the population size. The red curve represents the mass distribution given by the IDE. The dilution rate $ D $ is 0.25 h$^{-1}$. Again we observe the convergence of the IBM solution to the IDE in large population limit.}
\label{fig.repartition.masse}
\end{figure}

Figures \ref{fig.evol.taille.concentrations} and  \ref{fig.repartition.masse} illustrate the convergence of IBM to EID. The variances in the evolutions of the biomass concentration and of the substrate concentration as well as the relative variance of the number of individuals decrease when we increase the number of individuals, see Figure \ref{fig.evol.taille.concentrations}. The normalized size distributions at times $t=1, \, 3$ and $80$ (h) are represented in Figure \ref{fig.repartition.masse} for the IDE (red curve) and 100 independent runs of the IBM (blue histograms) for the small, the medium and the large population. Note that the number of bins was adapted according to the scale of the population in order to obtain clear graphics.

\begin{figure}
\begin{center}
\includegraphics[trim=4.3cm 1.5cm 2.2cm 2.2cm,width=10cm]{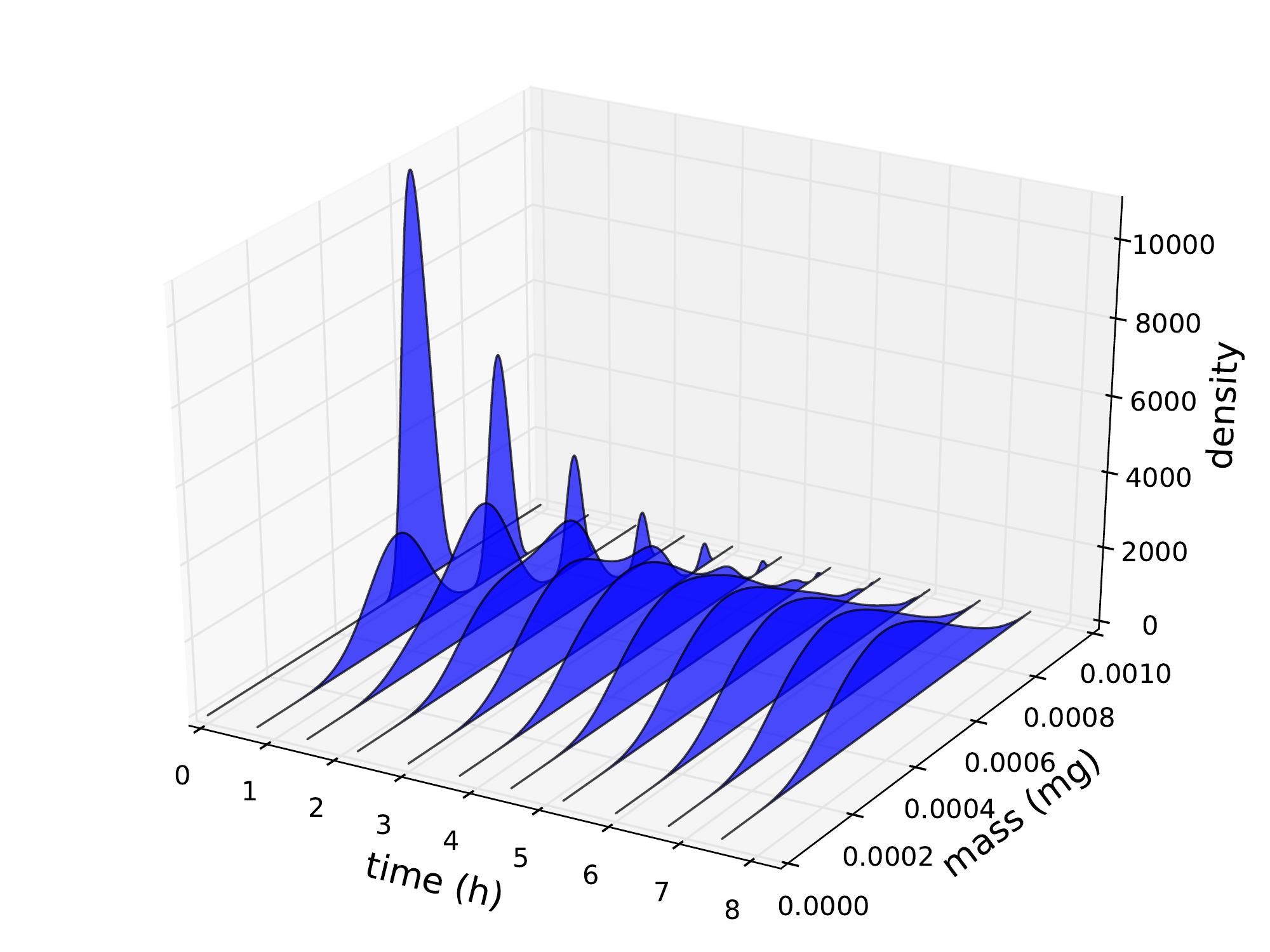}
\end{center}
\caption{Time evolution of the normalized mass distribution for the IDE \eqref{eq.limite.eid.fort}: we represent the simulation until time $T=8$ (h) only to illustrate the transient phenomenon caused by the choice of the initial distribution \eqref{eq.d}. After a few iterations in time this distribution is bimodal, the upper mode growths in mass and disappears before $T=8$ (h).}
\label{fig.evol.eid}
\end{figure}

The normalized solution of the IDE \eqref{eq.limite.eid.fort} is represented in Figure \ref{fig.evol.eid}. It corresponds to the time evolution of the normalized mass distribution. At the initial instant this distribution is given by the function \eqref{eq.d}. Then it becomes bimodal. The lower mode corresponds to the bacteria from the division. The upper mode represents bacteria of the initial distribution before their division or up-take. We observe the same phenomenon in the realization of IBMs, see Figure \ref{fig.repartition.masse}.
In contrast, the classic chemostat model presented below, see Equations \eqref{eq.chemostat.edo.1}-\eqref{eq.chemostat.edo.2}, cannot account for this phenomenon.
After this transient phenomenon, the normalized mass distribution converges to a stationary state.

As the IDE is the limit of the IBM in large population size, the behavior of the IDE gives informations on the behavior of the IBM. But there is no reason that the IDE corresponds to the mean value of the IBM, because of the correlation between the individuals behaviors.

\subsection{Comparison of the IBM, the IDE and the ODE}

We now compare the IBM and the IDE to the classic chemostat model described by the system of ODE's \eqref{eq.chemostat.edo.1}-\eqref{eq.chemostat.edo.2}.  The growth model in both the IBM and the IDE is of Monod type, so for the ODE model we also consider the classic Monod kinetics \eqref{eq.edo.monod}.
The parameters of this Monod law are not given in the initial model and we use 
a  least squares method to determine the value of the parameters  $\mumax$ and $\Ks$
which minimize the weighted quadratic distance between $(S_{t},X_{t})_{t\leq T}$ given by \eqref{eq.chemostat.edo.1}-\eqref{eq.chemostat.edo.2} and  $(\bar S_{t},\bar X_{t})_{t\leq T}$, where $\bar S_{t}$ and $\bar X_{t}$ are the means of the variable $S_t$ and $X_t=V^{-1}\,\int_{\X} x\,\nu_{t}(x)\,\rmd x$ given by the IBM \eqref{eq.xi}. This quadratic distance is weighted by the variance of the IBM.

\bigskip

\begin{figure}
\begin{center}
\includegraphics[width=14cm]{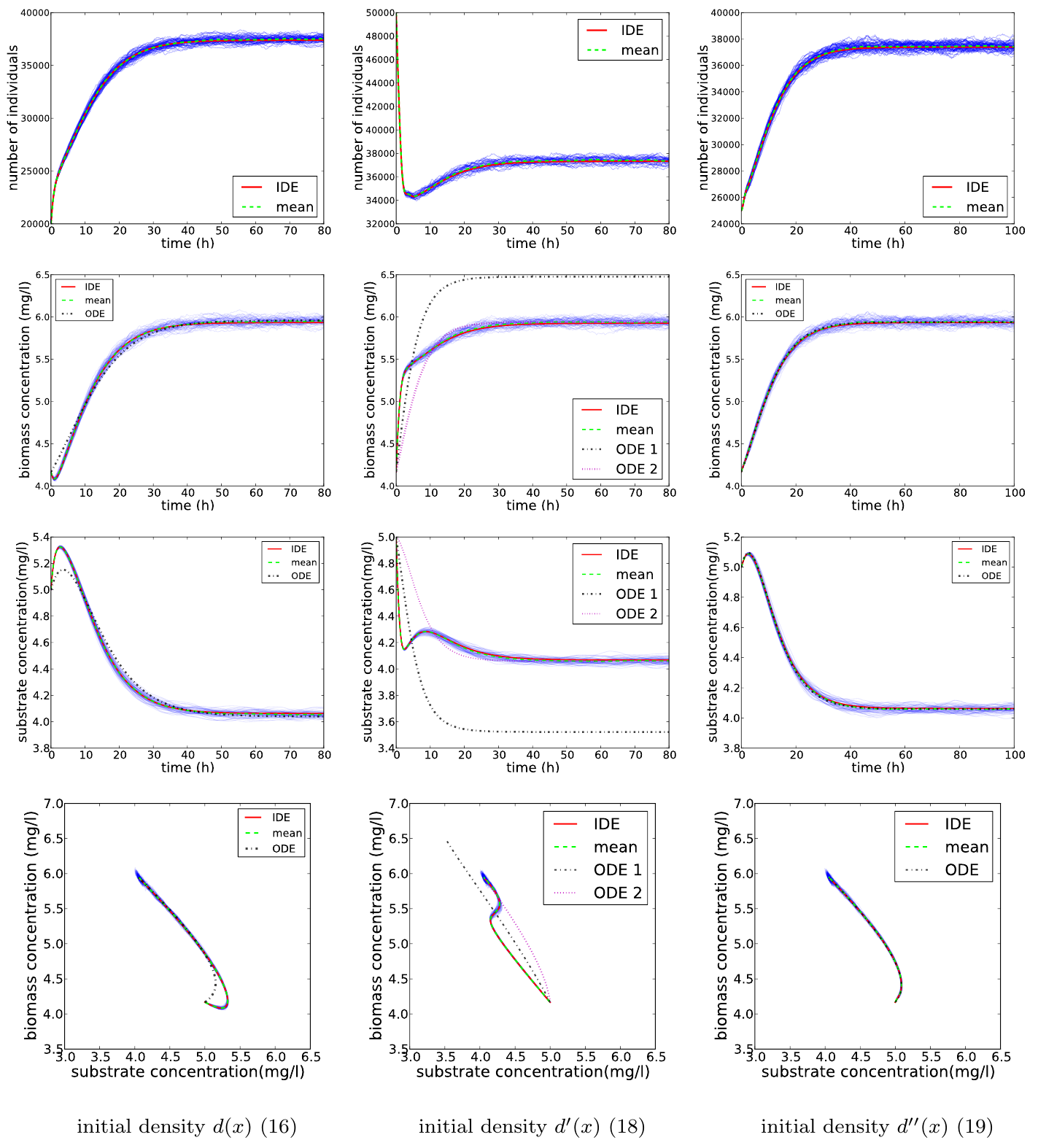}
\end{center}
\caption{Top to bottom : Time evolution of the number of individuals, the biomass concentration, the substrate concentration and the concentration trajectories in the phase space according to the initial mass distributions \eqref{eq.d} (left), \eqref{eq.d'} (middle) and \eqref{eq.d''} (right). In blue, the trajectories of 60 independent runs of the IBM simulated with $V=3$ l and $N_0=20000$ (left), $N_0=50000$ and $N_0=25000$ (right); in green, the mean of the IBM runs; in red, the solution of IDE \eqref{eq.limite.substrat.fort}-\eqref{eq.limite.eid.fort}; in black, the solution of the system \eqref{eq.chemostat.edo.1}-\eqref{eq.chemostat.edo.2}. The latter is fitted by the least squares method on the IBM, the parameters of the Monod law \eqref{eq.edo.monod} are $\mu_{\max}=0.329$ and $K_s = 2.603$ (left), $\mu_{\max}=11.556$ and $K_s = 200.0$ for EDO1 and $\mu_{\max}=9.219$ and $K_s = 183.065$ for EDO2 (middle), $\mu_{\max}=0.397$ and $K_s = 3.991$ (right). 
Note that in the case \eqref{eq.d'} (middle) it is unrealistic to fit a classic EDO chemostat to the IBM, indeed it is not possible for that model to fit the transitory behavior of the IBM leading to unrealistic values for the parameters of the chemostat.
}
\label{fig.edo.ibm.eid}
\end{figure}

Figure \ref{fig.edo.ibm.eid} represents evolution of the number of individuals, the biomass concentration, the substrate concentration and the trajectories in the phase space for  60 independent runs of the IBM and for the IDE with parameters of the Table \ref{table.parametres2} and with different initial density. The initial number $N_{0}$ is adapted so that the average initial biomass concentration is the same in the three cases.

First we consider  a simulation based on the initial mass density $d(x)$  defined by 
\eqref{eq.d}. With this initial density both the IDE and the IBM feature a transient phenomenon described in the previous section and illustrated in Figures  \ref{fig.evol.eid} and~\ref{fig.repartition.masse}. 
Figure \ref{fig.edo.ibm.eid} (left) shows a significant difference between the IBM and the IDE on the one hand and the ODE on the other hand, the latter model cannot account for the transient phenomenon. With the first two models, the individual bacteria are withdrawn uniformly and independently of their mass (large mass bacteria has the same probability of withdrawal as small mass bacteria).
As the initial state $d(x)$ has a substantial proportion of large bacteria mass, we have an important division rate at the population scale and a relatively low growth of individual (see Figure \ref{fig.parameters}). Therefore at the beginning of the simulation there is an important increase of the number of individuals whereas the biomass decrease. The ODE is naturally not able to account for this transient phenomenon.

Conversely, if we choose an initial density which charges the low masses, as the following
\begin{align}
\label{eq.d'}
d'(x)
	& =	
	    \frac{1}{C_{d'}} \,
		\Biggl(
			\frac{x-0.000125}{0.00025}
			\,\left(1-\frac{x-0.000125}{0.00025}\right)
		\Biggr)^5 \,
		1_{\{0.000125 < x < 0.000375\}}\,
\end{align}
where $C_{d'}$ is a normalizing constant, we observe an important increase of the biomass at the beginning of the simulation for the IBM and the IDE whereas the number of individuals decrease (see Figure \ref{fig.edo.ibm.eid} (middle)), which is due to fact that at the beginning of the simulation individuals have masses too low to divide, but with a high ``speed of growth'' (see Figure \ref{fig.parameters}).
As the randomness is low at the beginning of the simulation of IBMs, the least squares method, weighted by the variance of IBMs, give an ODE which have a strong increase of the biomass concentration and a strong decrease of the substrate concentration near the initial instant, but the stationary state of the ODE (black curves) doesn't match to the quasi-stationary state of the IBM or the stationary state of the IDE. 
If we give a high weight to the quasi-stationary state (between $t=40$ and $t=80$), we obtain an ODE (magenta curves) with a stationary state which matches to the quasi-stationary state of the IBM, but with a strong difference during the transitory state.

This phenomenons no longer appear if we use the following density:
\begin{align}
\label{eq.d''}
d''(x)
	& =	
	    \frac{1}{C_{d''}} \,
		\Biggl(
			\frac{x-0.00035}{0.0003}
			\,\left(1-\frac{x-0.00035}{0.0003}\right)
		\Biggr)^5 \,
		1_{\{0.00035 < x < 0.00065\}}\,.
\end{align}
where $C_{d''}$ is a normalizing constant.
Indeed, from Figure \ref{fig.edo.ibm.eid} (right), the different simulations are comparable, the ODE and the IDE match substantially.

\begin{table}[h]
\begin{center}
\begin{tabular}{|c|c|}
	\hline
    Parameters & Values \\
    \hline
    $S_0$				&	5 mg/l \\
    $\Sin$				&	10 mg/l \\
    $D$					& 	0.2 h$^{-1}$\\
    $\mmax$				&	0.001 mg \\	
    $\mdiv$				&	0.0004 mg \\
	$\bar\lambda$		&	1 h$^{-1}$\\
    $p_\lambda$			&	1000 \\
    $p_\beta$			&	7 \\
    $\rmax$				&	1 h$^{-1}$\\
    $k_r$				&	10 mg/l\\
    $k$					&	1\\
    \hline
\end{tabular}
\end{center}
\caption{Simulation parameters.}
\label{table.parametres2}
\end{table}

\bigskip

\begin{figure}
\begin{center}
\begin{tabular}{cc}
\includegraphics[width=7cm]{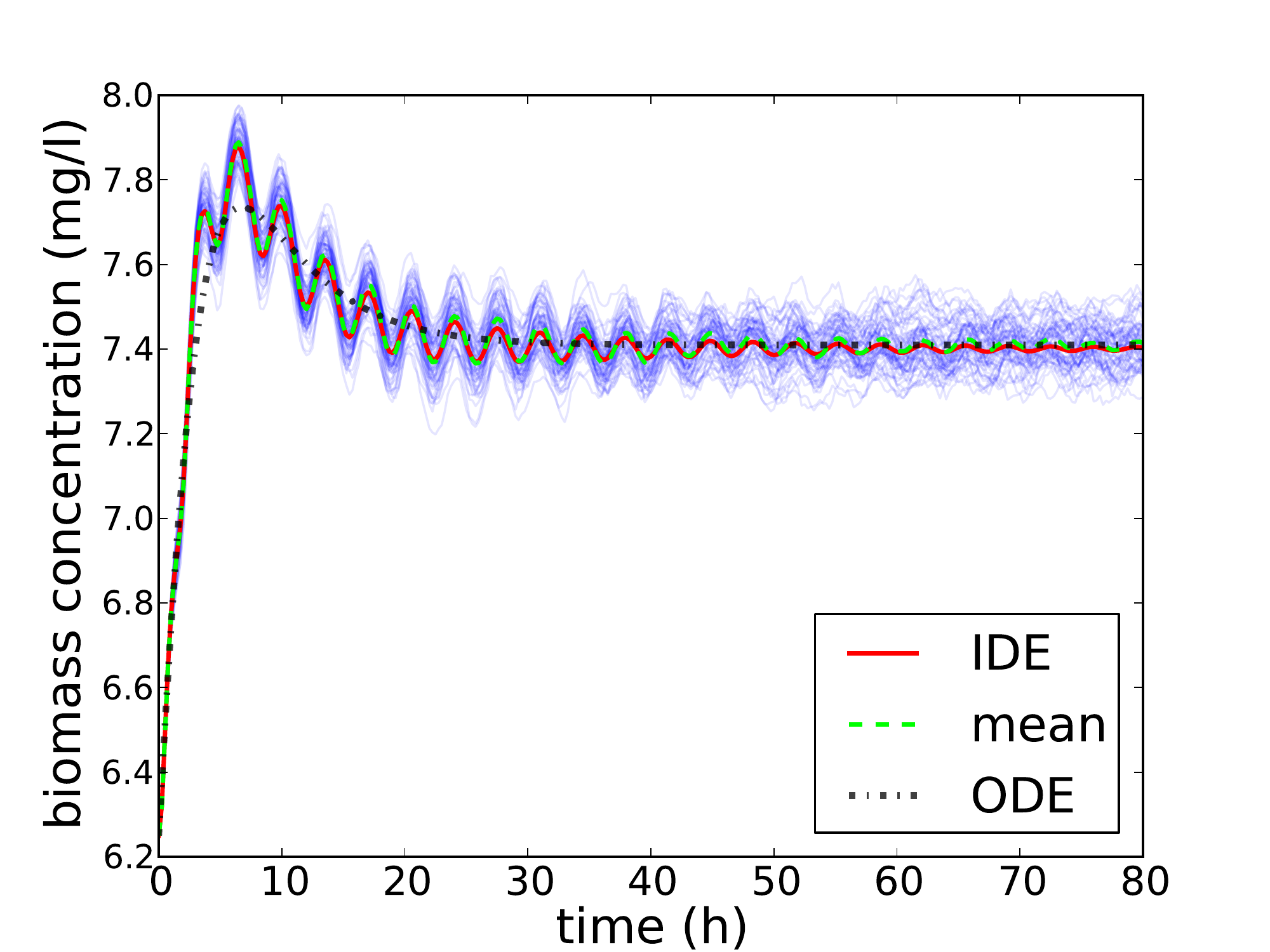}
&
\includegraphics[width=7cm]{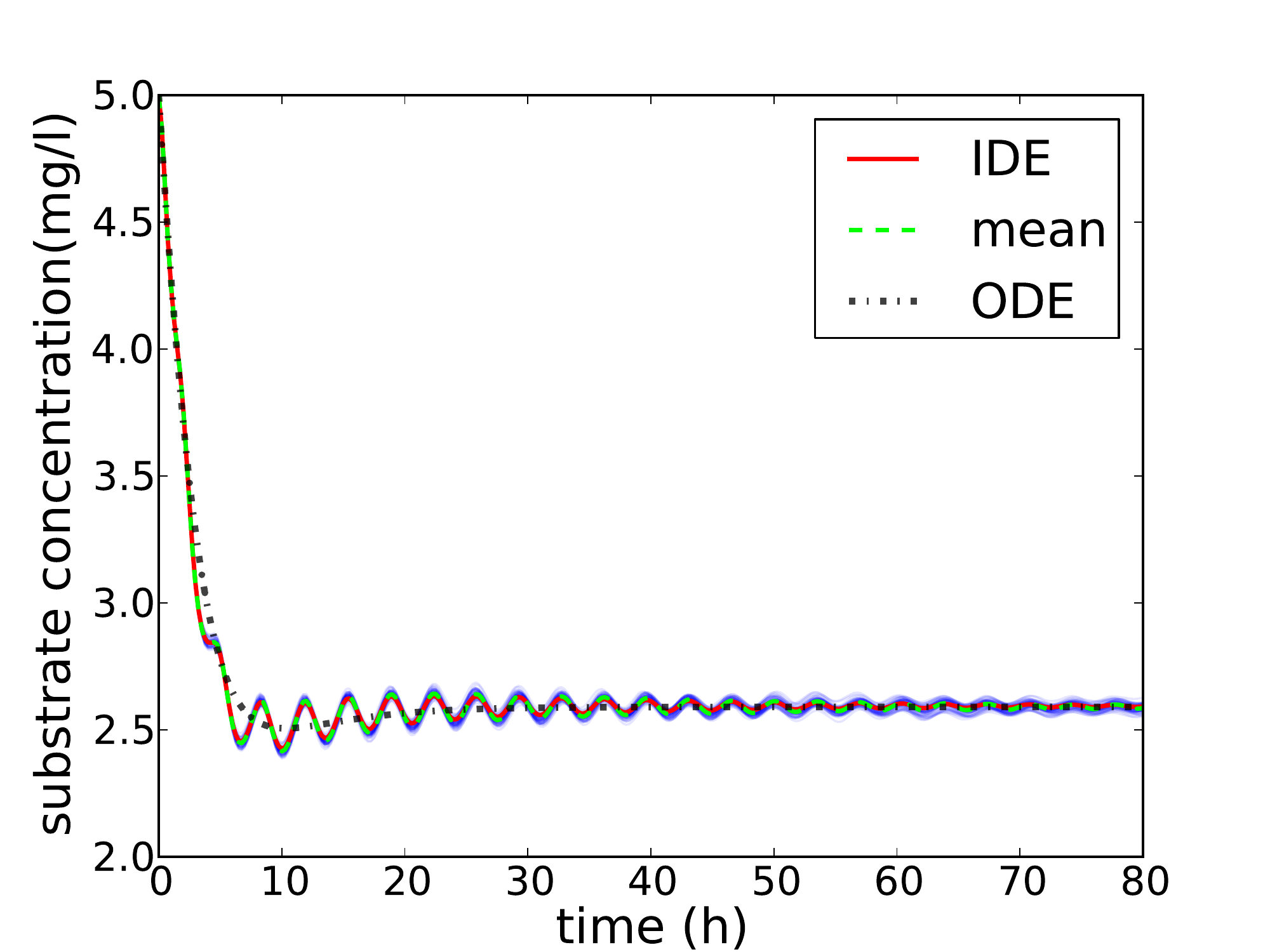}
\end{tabular}
\end{center}
\caption{Evolutions of the biomass (left) and the substrate (right) concentrations of 60 independent runs of the IBM (blue), the mean of the IBM (green), the IDE (red), the ODE (black) fitted by the least squares method on the IBM. The parameters of the Monod law \eqref{eq.edo.monod} of the ODE are $\mu_{\max}=0.537$ and $K_s=4.363$. The division rate function is given by the equation \eqref{eq.lambda2}. $\bar\lambda=5$ h$^{-1}$, $\mdiv=0.0005$ mg, $p_\beta=100$, $V=1.0$ l, $N_0=10000$. Other parameters are given in the Table \ref{table.parametres2}.}
\label{fig.edo.ibm.eid.oscillations}
\end{figure}

Figure \ref{fig.edo.ibm.eid.oscillations} shows simulations with the following division rate function :
\begin{align}
\lambda(s,x) = \bar\lambda \, 1_{\{x \geq \mdiv \}}, 
\label{eq.lambda2}
\end{align}
with $\bar\lambda=5$ h$^{-1}$, $\mdiv=0.0005$ mg and the parameter of the division kernel is $p_\beta=100$.

\begin{figure}
\begin{center}
\begin{tabular}{cc}
\includegraphics[trim=4.3cm 1.5cm 2.2cm 2.2cm,width=8cm]{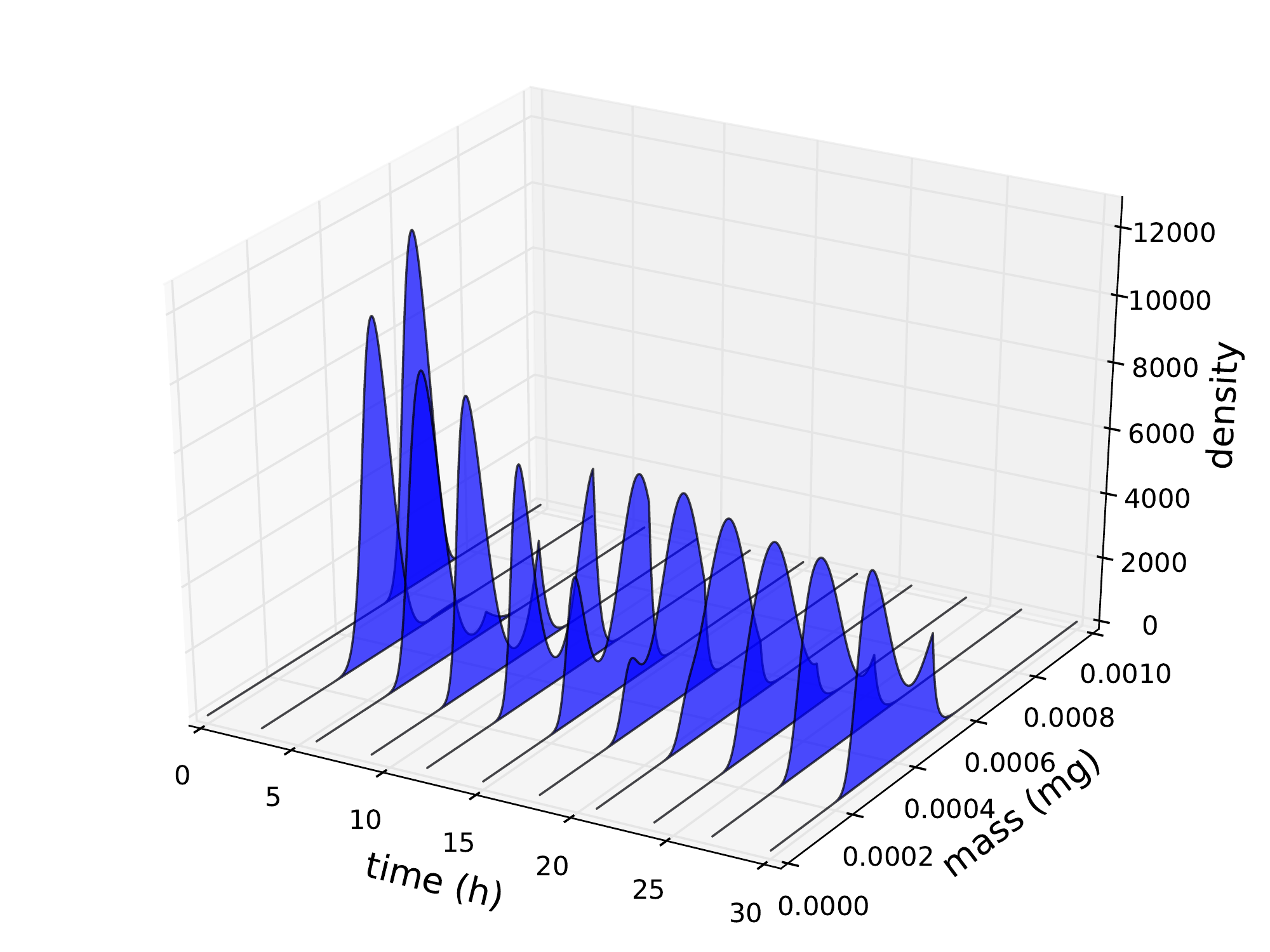}
\end{tabular}
\end{center}
\caption{Time evolution of the normalized mass distribution for the IDE \eqref{eq.limite.eid.fort} with the division rate function \eqref{eq.lambda2}, $\bar\lambda=5$ h$^{-1}$, $\mdiv=0.0005$ mg, $p_\beta=100$, $V=1.0$ l, $N_0=10000$. Other parameters are given in the Table \ref{table.parametres2}.}
\label{fig.evol.eid.oscillations}
\end{figure}

Another interesting  phenomenon is that we 
 can observe oscillations in the evolutions of the biomass and substrate concentrations for the IBM and the IDE, which can not be accounted by the ODE. This oscillations are due to the distribution which stay bimodal with alternation of the higher density between the lower and the upper mode (see Figure \ref{fig.evol.eid.oscillations}). When the lower mode have a higher density than the upper mode, there are a lot of individuals which quickly grow, then the biomass  concentration increases and the substrate concentration decreases. When the upper mode has a higher density than the lower mode, there are more individuals with a low growth, then the biomass concentration decreases and the substrate concentration increases.

\subsection{Study of the washout}

\begin{figure}
\begin{center}
\includegraphics[width=10cm]{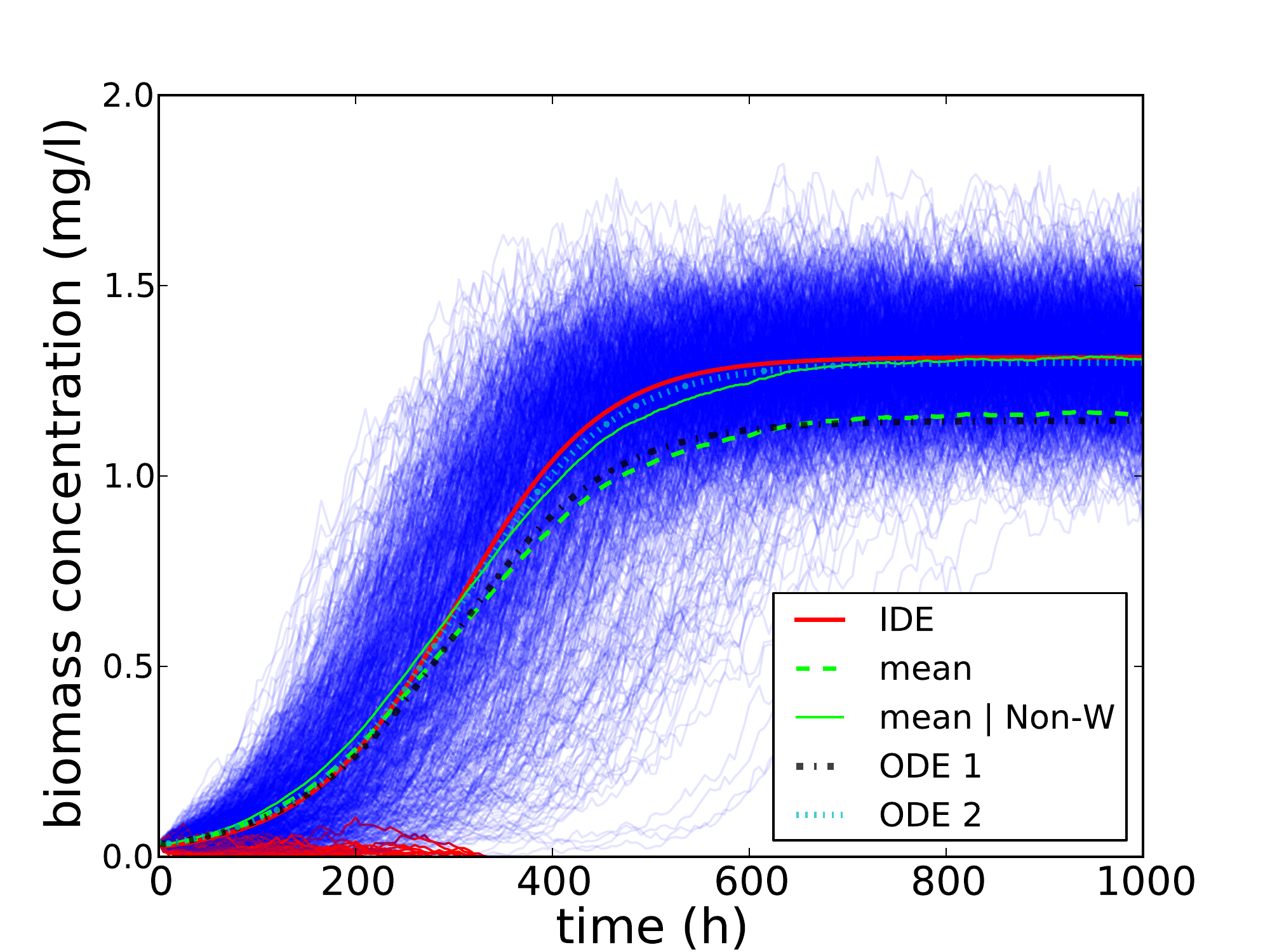}
\end{center}
\caption{Time evolution of the biomass concentration. In blue, 1000 independent realizations of the IBM simulated with $V=0.5$ l and $N_0=30$;  in green, the mean of these runs; in red, the solution of the IDE; in black, the solution of the ODE 1 with parameters values $\mumax=0.432$ and $\Ks = 5.050$, fitted on the IBM, weighted by the variance. In cyan, the solution of the ODE 2 with parameters values $\mumax=0.406$ and $\Ks = 4.142$, fitted on the IBM given by the non extinction of the population, weighted by the variance of non-extinct populations. Parameters are given by the Table \ref{table.parametres2}. The dilution rate $ D $ is   0.275 h$^{-1}$.  Among the 1000 independent runs of the IBM, 111 lead to washout while the deterministic models converge to an equilibrium with strictly positive biomass. The mean value of the  1000 runs of the IBM gives account for the washout probability while IDE and ODE models do not account for this question.}
\label{fig.lessivage}
\end{figure}

\begin{figure}[p]
\begin{center}
\includegraphics[width=9cm]{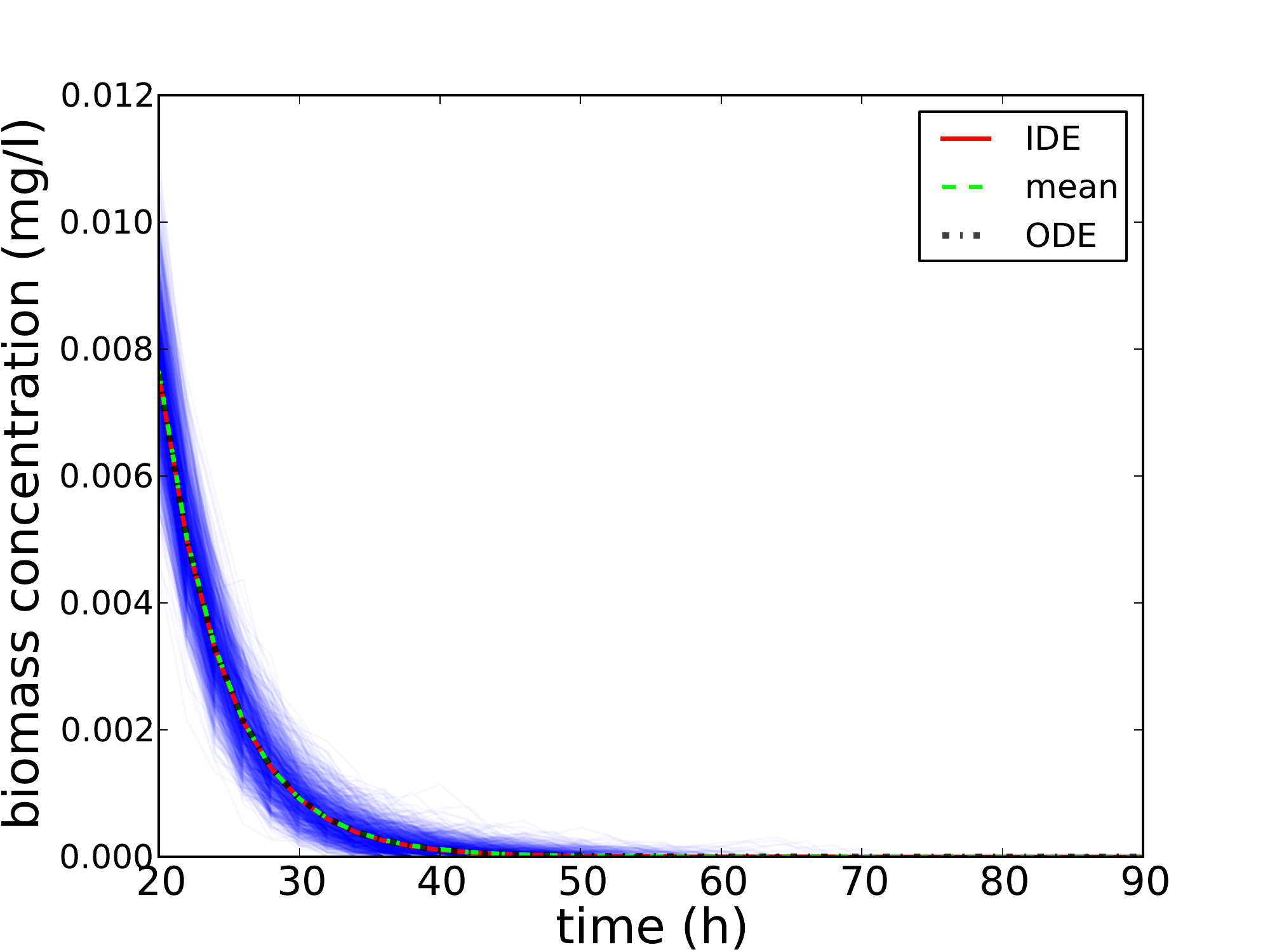}\\
\includegraphics[width=9cm]{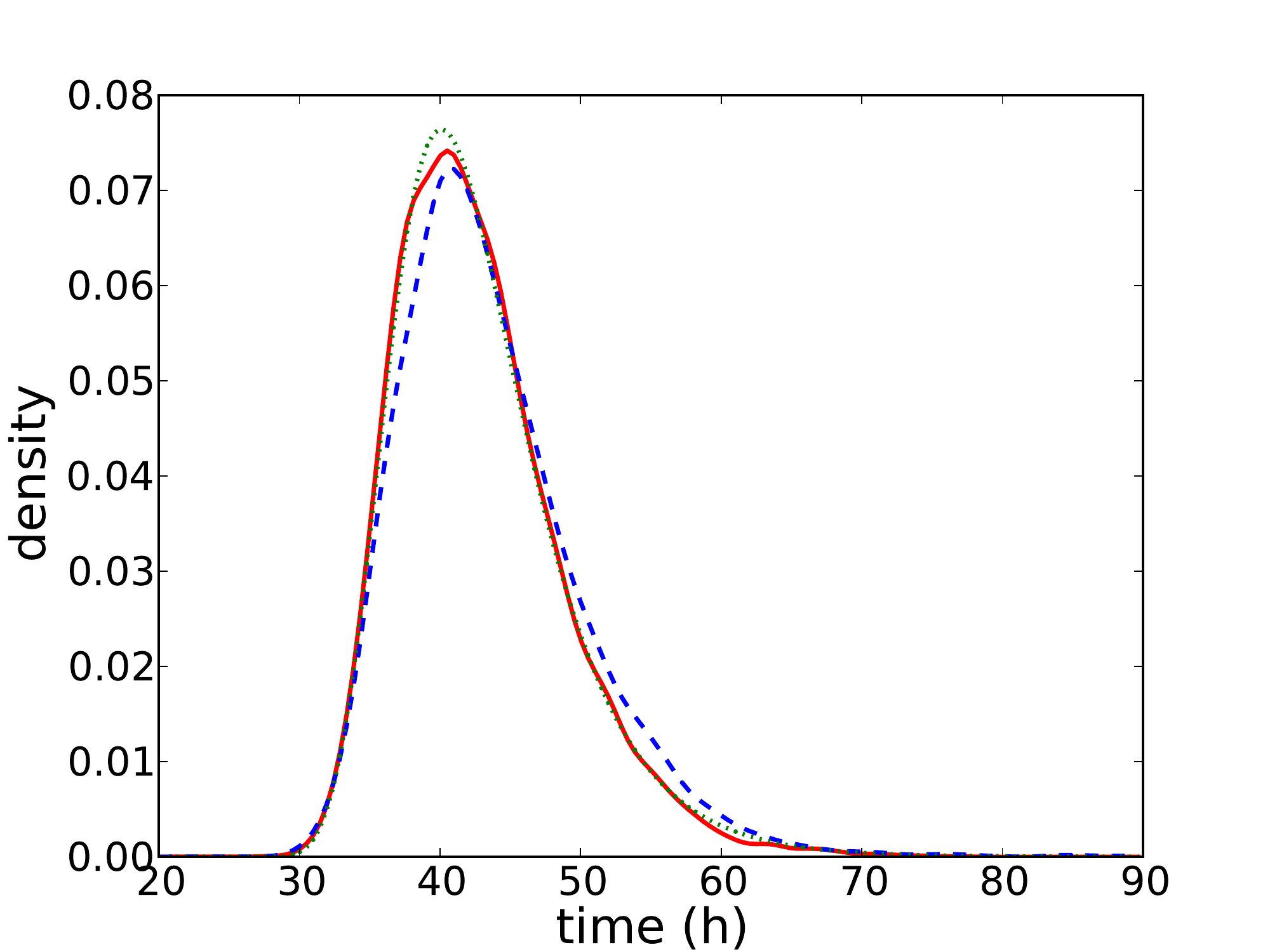}
\end{center}
\caption{$\blacktriangleright$ (Top) Evolution of biomass concentration between $t=20$ and $t=90$ h: blue, 1000 independent runs of the IBM; in green, the mean value of these runs; in red the solution of the IDE; in black, the solution to the ODE with parameters $\mu_{\max}=0.578$ and $K_s = 10.0$. The parameters are $V=10$ l and $N_0=10000$, the dilution rate $D$ is 0.5 h$^{-1}$, others parameters are the ones of the Table \ref{table.parametres2}. For both deterministic models, the size of the population decreases exponentially rapidly to 0 but remains strictly positive for any finite time. However, all the runs of the IBM reach washout in finite time. 
$\blacktriangleright$~(Bottom) The continuous red line is empirical distribution of the washout time calculated from 7000 independent runs of the IBM and plotted using a time kernel regularization. 
The dashed blue line is the empirical distribution of the washout time calculated from 7000 independent runs of the birth-death process with the same parameters as the ODE matched on the IBMs. The distribution is also plotted using a time kernel regularization.
The green dotted line is the p.d.f. \eqref{pdf.extinction.time} with $N_0=10000$, $D=0.5$ et $\tilde\lambda=0.2922$.}
\label{fig.lessivage2}
\end{figure}

One of the main differences between deterministic  and stochastic models lies in their way of accounting for the washout phenomenon  (or extinction phenomenon in the case of an ecosystem). With a sufficiently small dilution rate $D$, the solutions of the system \eqref{eq.chemostat.edo.1}-\eqref{eq.chemostat.edo.2} and of the IDE  \eqref{eq.limite.substrat.fort}-\eqref{eq.limite.eid.fort}  converge to an equilibrium point with strictly positive biomass. In fact, the  washout is an unstable equilibrium point and apart from the line corresponding to the null biomass, the complete phase space corresponds to a basin of attraction leading to a solution with a strictly positive biomass asymptotic point. However, from Figure \ref{fig.lessivage}, among the 1000 independent runs of the IBM, 111  converge to washout before time $t=1000$ h; so the probability of washout at this instant is approximately 11\%. 
The ODE 1 (dot-dashed black line) is fitted to the 1000 IBMs. We can observe that it matches to the mean. 
The ODE 2 (dotted cyan line) is fitted on the non-extinct IBM and matches to the mean conditionally to the non extinction.   
It may be noted that the IDE and the ODE do not correspond to the average value of the IBM since only the latter may reflect the washout in a finite time horizon.

Now we consider a sufficiently large dilution rate, $D=0.5$ h$^{-1}$, corresponding to the washout conditions. Figure \ref{fig.lessivage2} (top) presents the evolution of the biomass concentration in the different models. The runs of the IBM converge to the washout in finite time whereas both deterministic ODE and IDE models converge exponentially to washout without ever reaching it in finite time. Figure \ref{fig.lessivage2} (bottom) shows the empirical distribution of the washout time calculated from 7000 independent runs of the IBM (red curve). This washout time features a relatively large variance.

It is known that for a birth-death process with constant rates $\tilde\lambda$ and $D$ which corresponds respectively to the rates of birth and death and with $\tilde\lambda<D$, the probability density function of the time of extinction $T$ is
\begin{align}
\label{pdf.extinction.time}
d(t)
	= &
		N_0 \, D \, 
		\frac{(\tilde\lambda-D)^2 \, e^{(\tilde\lambda-D)\,t}}
		{(\tilde\lambda \, e^{(\tilde\lambda-D)\,t}-D)^2} \,
		\left( 
			\frac{D \, e^{(\tilde\lambda-D)\,t}-D}
			{\tilde\lambda \, e^{(\tilde\lambda-D)\,t}-D}
		\right)^{N_0-1}.
\end{align}

When the birth rate is not constant, we can expect that the probability density function of the time of extinction is of the form \eqref{pdf.extinction.time} where $\tilde\lambda$ is the average birth rate of the population. 
Figure \ref{fig.lessivage2} shows the probability density function \eqref{pdf.extinction.time} (green dotted curve) where $\tilde\lambda$ is computed by a least squared method in order to be fitted on the empirical distribution of the washout time (red solid curve). 
This constant $\tilde \lambda$ depends on the model parameters, in particular on the initial number of individuals $N_0$ and on the initial distribution of individuals. In our exemple the initial distribution contains bacteria with higher masses than the quasi-stationary distribution, then the effective division rate near the time $t=0$ is higher than the quasi-stationary effective division rate and therefore, the constant $\tilde\lambda$ will be higher too. Moreover, higher the initial number of individuals $N_0$ is, more negligible the time the reach the quasi-stationary distribution is.
The dashed blue curve represents the empirical law of the extinction time of the BDP, calculated from 7000 independent runs of the BDP, where the function $\mu$ in equations \eqref{eq.bdp.s}-\eqref{eq.bdp.y} is a Monod function \eqref{eq.edo.monod} with the same parameters as the ODE fitted on the IBM.

\section{Discussion}
\label{sec.discussion}

In this work we presented four models of the chemostat together with the analytical and algorithmic gateways bridging one to the other:
\begin{align*}
\small
\begin{array}{ccccc}
\rotatebox[origin=c]{45}
{
\begin{minipage}{1.5cm}\centering\tiny deterministic \\ models\end{minipage}
}
&
\textrm{\fbox{ODE model \eqref{eq.chemostat.edo.1}-\eqref{eq.chemostat.edo.2}}} 
  & 
  \xleftarrow{\textrm{\tiny model reduction}}
  & \textrm{\fbox{IDE model \eqref{eq.limite.substrat.fort}-\eqref{eq.limite.eid.fort}}}
&
\hbox{\begin{minipage}{3cm}\raggedright\tiny 
classic numerical methods for ordinary differential equations and integro-differential equations\end{minipage}}
\\[-1em]
&
  \rotatebox[origin=c]{90}
     {$\xrightarrow[\textrm{\tiny population size}]{\textrm{\tiny large }}$}
  &
  &
  \rotatebox[origin=c]{90}
     {$\xrightarrow[\textrm{\tiny population size}]{\textrm{\tiny large }}$}
\\[-1em]
\rotatebox[origin=c]{45}
{
\begin{minipage}{1.5cm}\centering\tiny stochastic\\ models\end{minipage}
}
&
\textrm{\fbox{BDP model \eqref{eq.bdp.s}-\eqref{eq.bdp.y}}} 
  & 
  \xleftarrow{\textrm{\tiny model reduction}}
  & \textrm{\fbox{IBM model}}
&
\hbox{\begin{minipage}{3cm}\raggedright\tiny 
hybrid Monte Carlo algorithms, see Algorithms \ref{algo.ssa} and \ref{algo.ibm}
\end{minipage}}
\\
&
\begin{minipage}{1.5cm}\centering\tiny unstructured \\ models\end{minipage}
&
&
\begin{minipage}{1.5cm}\centering\tiny structured \\ models\end{minipage}
\end{array}
\end{align*}
On the one hand we considered the classic deterministic model of chemostat as a system of ODE's, and also a birth and death stochastic process hybridized with an ODE; on the other hand their mass-structured counterparts, a deterministic IDE and also a stochastic IBM hybridized with an ODE. In all cases the evolution of the substrat is represented as an ODE meaning that this part of the model is reasonably represented as a fluid limit dynamic.
The stochastic model are Markov processes with values in $\R_{+}\times\N$ for the unstructured model and with values in $\R_{+}\times \MM([0,\mmax])$ for the mass-structured model. The Markov property allows to analytically prove the convergence of the stochastic models toward their deterministic counterpart in large population size limit. Moreover the reduction from the mass-structured models to the unstructured ones is obtained by a simplification of the growth function.  

The numerical simulations of deterministic models are straightforward and are done using classic integration schemes. The numerical simulation of random models uses almost exact Monte Carlo algorithms, indeed the models are hybrid and the integration of the ODE part of the model is achieved through approximation schemes. These latter algorithms are not realistic in large population as all events, cell division and cell  uptake, are explicitly simulated; but it is precisely at this level that the simulation of the deterministic models took over, the whole framework being perfectly consistent.

It is important to evaluate the complexity of the models  in terms of analysis as well as simulation. For example, it is difficult to determine an \emph{optimal} control law for the IBM while this task is relatively easy in the case of the classic ODE model. In this latter case there is already a large number of results, while in the former case the criteria to optimize are still not well established. However, it is pertinent to test a control law developed on the ODE model \eqref{eq.chemostat.edo.1}-\eqref{eq.chemostat.edo.2} not on the same model but on simulated data generated from the~IBM.

\bigskip

Despite their relative complexity, stochastic discrete models are essential in more than one respect for population dynamics. On the one hand they allow to explore situations where deterministic models are totally blind, this is particularly the case for situations close to extinction conditions or near wash-out conditions in the case of chemostat. This question may also be relevant in larger population size \citep{campillo2012b}. On the other hand they offer a non-reproducible simulation tool close to conditions encountered in practice. As the biologist Georgy Gause already pointed out in 1934:
``When the microcosm approaches the natural conditions [...] the struggle for existence begins to be controlled by such a multiplicity of causes that we are unable to predict exactly the course of development of each individual microcosm. From the language of rational differential equations we are compelled to pass on to the language of probabilities, and there is no doubt that the corresponding mathematical theory of the struggle for existence may be developed in these terms'' \citep{gause1934a}.

However, the stochastic and discrete modeling is essentially devoted to evolutionary  population dynamics. It is only more recently that this approach is extended to all areas of population dynamics with a similar concern to encourage cooperation between different representations of a model \citep{andrews2009a}. It is interesting to note that the same approach is now also adopted in epidemiology where  considerations of discrete and random aspects of population dynamics in small sizes are essential \citep{allen2012a,allen2013a}.

The IBM proposed here is certainly not the most efficient in terms of computational speed: it is asynchronous and requires the simulation of each individual event. 
There are strategies that accelerate this IBM thanks to some approximations. The proposed IBM has the advantage of being an exact Monte Carlo simulation, up to approximation schemes of the ODE, of the very stochastic process which we can analyze and prove the  weak convergence in large population toward the ID model. This important property is due to the fact that all the models considered here, including the deterministic ones, are of Markov and that the study of weak convergence of these processes is an important tool in terms of mathematics but also on a practical level in terms simulation. 

Finally, this work advocates for the development of hybrid models relevant when the size of a given population fluctuates between large and small values, or when multiple populations are involved some in large sizes, others in small sizes.

\section*{Acknowledgements}

The authors are grateful to  Claude Lobry for discussions on the model, to Pierre Pudlo and Pascal Neveu for their help concerning the programming of the IBM. This work is partially supported by the project ``Modèles Numériques pour les écosystèmes Microbiens'' of the French National Network of Complex Systems (RNSC call 2012). The work of Coralie Fritsch is partially supported by the Meta-omics of Microbial Ecosystems (MEM) metaprogram of INRA.

\appendix
\section*{Appendices}
\addcontentsline{toc}{section}{Appendices}

\section{Numerical integration scheme for the IDE}
\label{appendix.schema.num}

To  numerically solve the system of integer-differential equations \eqref{eq.limite.substrat.fort}-\eqref{eq.limite.eid.fort}, 
we make use of finite difference schemes.

Given a time step $\Delta t$ and a mass step  $\Delta x = L/I$, with $I \in \NN^*$,
we discretize the time and mass space with:
\begin{align*}
	t_n & = n \, \Delta t \,
	&
	x_i & = i \, \Delta x\,.
\end{align*}
We introduce the following approximations:
\begin{align*}
	p_{n,i} & \simeq p_{t_n}(x_i) \,,
	&
	s_n & \simeq S_{t_n}\,.
\end{align*}
We also suppose first that at the initial time step there is no individual with null mass in the vessel, i.e. $p_{0,0}=0$; and second that 
individual with null mass cannot be generated during the cell division step,
i.e.  $q$ is regular with $q(0)=0$. This assumption was not necessary in the mathematical development presented in the previous sections but is naturally required to obtain reasonable mass of individuals in the simulation. 

\medskip

For time integration we use an explicit Euler scheme, for space integration,
an uncentered upwind difference scheme, which leads to the coupled integration scheme:
\begin{align*}
	\frac{p_{n+1,i}-p_{n,i}}{\Delta t}
	&  =  
		- \rho(s_n,x_i)\, \frac{p_{n,i}-p_{n,i-1}}{\Delta x}	
		- \frac{\partial}{\partial x}\rho(s_n,x_i)\,p_{n,i} 
\\	
	&\qquad\qquad
		- \bigl(\lambda(s_n, x_i)+D \bigr)\,p_{n, i} 
	 	+ 2\, \Delta x \, \sum_{j=1}^{I}
	 		 				\frac{\lambda(s_n, x_j)}{x_j}\,
	 		 				q\left(\frac{x_i}{x_j} \right)\,
	 		 				p_{n,j} \,,
\\
	\frac{s_{n+1}-s_n}{\Delta t}
	& = 
	D\,(\Sin - s_n)
	- \frac{k}{V} \, \Delta x \, \sum_{j=1}^{I} \rho(s_n, x_j) \, p_{n,j}
\end{align*}
for $n \in \NN$ and $i = 1, \cdots I$, with the boundary condition:
\begin{align*}
	p_{n+1,0} = 0
\end{align*}
and given initial conditions $p_{0,i}$ and $s_{0}$.

We finally get:
\begin{align*}
	p_{n+1,i}
	& =   
 	p_{n,i} + \Delta t \;
	\Biggl\{
		- \rho(s_n,x_i)\, \frac{p_{n,i}-p_{n,i-1}}{\Delta x}	
		- \frac{\partial}{\partial x}\rho(s_n,x_i)\,p_{n,i} 
\\	
	& \qquad\qquad
		- \bigl(\lambda(s_n, x_i)+D \bigr) \, p_{n, i} 
	 	+ 2\, \Delta x \, 
			\sum_{j=1}^{I} \frac{\lambda(s_n, x_j)}{x_j}\,
	 		 			q\left(\frac{x_i}{x_j} \right)\, p_{n,j} 
		\Biggr\}
\\
	s_{n+1}
	& = 
	s_n + \Delta t \;  
	\Biggl\{
		D\,(\Sin - s_n) 
		- \frac{k}{V} \, \Delta x \, \sum_{j=1}^{I} \rho(s_n, x_j) \, p_{n,j}
	\Biggl\}
\end{align*}
for  $n \in \NN$ and $i = 1, \cdots I$ with boundary condition  $p_{n+1,0} = 0$ and given initial conditions $p_{0,i}$ and $s_{0}$.



\end{document}